\RequirePackage{fixltx2e}
\documentclass[preprint, iop]{emulateapj}
\usepackage{comment}
\pdfoutput=1
\usepackage{amsmath}
\usepackage{longtable}
\usepackage[breaklinks,colorlinks,urlcolor=cyan,citecolor=cyan,linkcolor=cyan]{hyperref}
\usepackage{url}

\shorttitle{Constraints on Planet Nine}
\shortauthors{Millholland and Laughlin}

\begin{document}

\title{Constraints on Planet Nine's Orbit and Sky Position within a Framework of Mean Motion Resonances}
\author{Sarah Millholland and Gregory Laughlin}
\affil{Department of Astronomy, Yale University, New Haven, CT, 06511, USA}
\email{Email: sarah.millholland@yale.edu}

\begin{abstract}
% Word limit: 250
A number of authors have proposed that the statistically significant orbital alignment of the most distant Kuiper Belt Objects (KBOs) is evidence of an as-yet undetected planet in the outer solar system, now referred to colloquially as ``Planet Nine''. %The consensus that such a planet may exist has been growing for several years, as the number of known aligned objects has increased. 
Dynamical simulations by \cite{2016AJ....151...22B} have provided constraints on the range of the planet's possible orbits and sky locations. We extend these investigations by exploring the suggestion of \cite{2016ApJ...824L..22M} that Planet Nine is in small integer ratio mean-motion resonances (MMRs) with several of the most distant KBOs. We show that the observed KBO semi-major axes present a set of commensurabilities with an unseen planet at $\sim 654$ AU ($P \sim 16,725$ yr) that has a greater than 98\% chance of stemming from a sequence of MMRs rather than from a random distribution. %Probably too aggressive a statement, but perhaps not that unwarranted...
We describe and implement a Monte-Carlo optimization scheme that drives billion-year dynamical integrations of the outer solar system to pinpoint the orbital properties of perturbers that are capable of maintaining the KBOs' apsidal alignment. This optimization exercise suggests that the unseen planet is most consistently represented with mass, $m \sim 6-12 M_{\oplus}$, semi-major axis, $a \sim 654$ AU, eccentricity, $e\sim0.45$, inclination, $i \sim 30^{\circ}$, argument of periastron, $\omega \sim 150^{\circ}$, longitude of ascending node, $\Omega \sim 50^{\circ}$, and mean anomaly, $M \sim 180^{\circ}$. A range of sky locations relative to this fiducial ephemeris are possible. We find that the region $30^{\circ}\, \lesssim \mathrm{RA} \lesssim 50^{\circ}$,
$-20^{\circ}\, \lesssim \mathrm{Dec} \lesssim 20^{\circ}$ is promising. 

%Within this area, a perturbing planet can simultaneously maintain long-term apsidal alignment among the known distant KBOs, while also exerting a dynamical influence that keeps several of the KBOs (in particular, Sedna, 2012 VP\textsubscript{113}, and 2010 GB\textsubscript{174}) librating in MMRs.

%A number of sky locations are plausible, but the largely unsurveyed region encompassed by $6.5\,{\rm h} \lesssim RA \lesssim 10 \,{\rm h}$ appears particularly promising. Finally, the simulations provide further evidence that MMRs are an important mechanism in the observed clustering of the distant KBO orbits. 

% Maybe not enough room for this:
%Consideration of resonance widths generates predicted bounds on the KBO semi-major axes. These bounds are substantially smaller than current measurement uncertainties. Repeated observations of these KBOs can therefore permit rapid testing of the predictions, to either refute or substantially strengthen the validity of our assumed scenario.
\end{abstract}

\section{Introduction}

The astronomical community has once again been captivated by the suggestion that there exists a massive, as-yet undetected trans-Neptunian planet. Interest has been particularly acute in response to the publication of an article by \cite{2016AJ....151...22B} that explicitly connects the spatial clustering of the perihelia of the most distant Kuiper Belt objects (KBOs) with an object, ``Planet Nine'', for which they estimate a mass $m\sim10\,M_{\oplus}$, a semi-major axis, $a\sim700\,$ AU, an eccentricity, $e\sim0.6$, and an orbital orientation that is anti-aligned with the roster of distant KBOs.

The prospect of discovering such a world is not new, and indeed, has been held out with varying intermittency and credibility for nearly 170 years.  The first quantitative prediction of a planet beyond Neptune was issued in 1848, shortly after the discovery of Neptune itself by Leverrier, Galle \& d'Arrest; the physicist Jacques Babinet analyzed discrepancies between the observed orbit of Neptune and the orbit initially derived by Leverrier. Babinet used the outcome of these calculations to posit the existence of a $m\sim13\,M_{\oplus}$ planet at a mean distance $a\sim47$ AU, a prediction that was, like most that followed, incorrect. Several decades later, in 1880, George Forbes was the first to propose the existence of trans-Neptunian planets that were somewhat similar to the Planet Nine model that is currently attracting interest. Forbes' outer planet had a semi-major axis $a\sim300$ AU, and drew its location from an analysis of the clustering of the aphelion distances of periodic comets. An early comparative photographic search for Forbes' predicted planets was carried out to a then-impressive depth of 15th magnitude, albeit with a null result that was reported in 1892 \citep{1980pxp..book.....H}. 

The long-running quest has recently achieved a particularly strong impetus with the realization that unusual properties of the most distant Kuiper Belt Objects can be ascribed to the dynamical influence of a sizeable perturbing body. \cite{2014Natur.507..471T} and \cite{2014MNRAS.443L..59D} observed that the distant scattered objects with $a \gtrsim 150$ AU and $q \gtrsim 30$ AU have arguments of perihelion clustered around $\sim 0^{\circ}$, and they considered a super-Earth at several hundred AU as a possible explanation. Recently, \cite{2016AJ....151...22B} observed that the most distant Kuiper Belt Objects are clustered not just in their arguments of perihelion but also in physical space. They showed that bodies on highly eccentric, nearly Neptune-crossing orbits can maintain clustering in their perihelia when they are apsidally anti-aligned with a massive, high eccentricity perturber with semi-major axis, $a \sim 700$ AU. Following this result, several authors elucidated further details connected to the proposed planet. Planet Nine might have a role in explaining the $6^{\circ}$ obliquity of the Sun \citep{2016AJ....152..126B, 2016arXiv160705111G, 2016arXiv160801421L} and the existence of the population of highly inclined, sometimes retrograde trans-Neptunian objects such as Drac and Niku \citep{2015Icar..258...37G, 2016AJ....151...22B, 2016ApJ...833L...3B, 2016ApJ...827L..24C}. 

Potential formation scenarios for such a distant and eccentric planet include scattering with or capture from nearby stars in the Sun's birth cluster \citep{2016ApJ...823L...3L, 2016MNRAS.460L.109M}, in situ accretion \citep{2016ApJ...825...33K}, or early scattering interactions with the giant planets \citep{2016ApJ...826...64B}. \cite{2016ApJ...824L..25F}, \cite{2016A&A...592A..86T}, and \cite{2016A&A...589A.134L} characterized various potential physical properties, such as the estimated size, the temperature, and the apparent brightness, which are useful when considering its detectability. 

The search space on the planet's current sky position has already been significantly constrained. \cite{2016ApJ...824L..23B} simulated a disk of scattered eccentric planetesimals subject to the influence of Planet Nine over a grid of potential orbital parameters. They searched for consistency with the properties of the observed KBO orbits and combined the resulting restricted parameter domain with the observational limits of past and present sky surveys. Others have exploited additional sources of data in effort to constrain the sky position. \cite{2016AJ....152...80H} used the several-decades-long ephemerides of Pluto and other trans-Neptunian objects to locate regions of the sky where the orbital fits are worsened or improved by the tidal acceleration from Planet Nine, thereby ruling out large areas of parameter space. \cite{2016A_A...587L...8F} and \cite{2016AJ....152...94H} derived similar constraints using the precise measurements of the Earth-Saturn distance obtained by the \textit{Cassini} spacecraft. \cite{2016AJ....152...94H} find a most favored location of RA $\sim 40^{\circ}$ and Dec $\sim -15^{\circ}$, extending $\sim 20^{\circ}$ in all directions. These results, however, are acknowledged to be very sensitively dependant on the precise solar system model that is employed to make a detailed fit to the observations (see e.g. \cite{2016DPS....4812007F}, and the press statement issued by JPL\footnote{\url{http://www.jpl.nasa.gov/news/news.php?feature=6200}}).

Irrespective of the constraints, the area of search space on the sky is still quite large. To make progress, it is useful to return to the details of the physical interaction between the proposed planet and the distant KBOs. \cite{2016AJ....151...22B} suggested that mean motion resonances (MMRs) may be the physical mechanism responsible for the clustering in the arguments of perihelion of the detached KBOs into a configuration that is anti-aligned to the orbit of Planet Nine. In Batygin and Brown's framework, the mean-motion commensurabilities protect the bodies from recurrent, destabilizing close encounters that would otherwise be expected for such highly eccentric, indeed crossing, orbits. Evidence that such phase protection may be at work was bolstered by the identification of test particles in their simulations stably librating in 2:1, 3:1, and other higher order MMRs for several hundred million years.  

Building on this evidence, \cite{2016ApJ...824L..22M} noticed that several of the detached KBOs could be in N:1 and N:2 MMRs if the semi-major axis of Planet Nine was $\sim$ 665 AU, corresponding to a 3:2 mean-motion resonance with Sedna. \citet{2016A_A...590L...2B} found consistent results using semi-analytical integrations of the resonant secular Hamiltonians, showing that high eccentricity, apsidally anti-aligned libration islands were possible in a resonant context. He also found, however, that similar libration islands exist in a scenario exhibiting purely secular dynamics, suggesting that MMRs are possible but are not strictly necessary to produce the observed clustering. 

Given these clues, we seek to further explore the conjecture that Planet Nine participates in MMRs with one or more detached KBOs. In this picture, Planet Nine's allowed semi-major axis is strongly constrained, and critically, participation in MMRs potentially permits determination of both Planet Nine's orbital properties, and its current sky position.  

Throughout the majority of this paper, we make the fundamental assumption that Planet Nine \textit{exists}. The results are thus best interpreted as the constraints on Planet Nine under this hypothesis. Towards the end of the paper, we adopt a contrasting viewpoint and comment on the implications of our results on the evidence for the planet's existence.

\begin{comment}
Our goal with this paper is spur Planet Nine's discovery by pinning down its sky position as accurately as possibility. In support of this objective, the plan is as follows: In \S2, we quantify the likelihood that the observed period ratios of the distant KBOs are suggestive of resonant relationships with an unseen perturber. In \S3, we describe the Monte-Carlo optimization scheme that we have used to evaluate the fitness of potential Planet Nine orbits and locations. In \S4, we present the results of the calculations, and assign likelihoods to specific regions of the sky, along with prediction of the likely visual and infrared appearance, and the sky motion of the planet given our models. In \S5, we compare our prospective sky locations with the results of both known searches, and existing optical and infrared catalogs, with a focus on eliminating particular swathes from consideration. In \S6, we give an overview of potential search strategies and conclude.
\end{comment}

The outline is as follows: In \S2, we construct a semi-major axis probability distribution under a resonant hypothesis and quantify the likelihood that the observed semi-major axes of the distant KBOs are suggestive of resonant relationships with Planet Nine. In \S3 we adopt the resonant hypothesis and describe the Monte-Carlo optimization scheme we have used to pinpoint the orbital elements of a planetary perturber that is capable of maintaining clustering in the KBO orbits to the degree that is observed today.  In \S4, we present the results of the calculations. We first examine the orbital element constraints and then demonstrate some key connections between maintenance of orbital clustering and participation in MMRs. In \S5, we compare our prospective sky locations with previous works and propose a plan to obtain stronger constraints. We also comment on the implications of our results on the evidence for Planet Nine's existence. We conclude in \S6.

\section{A semi-major axis probability distribution} \label{section2}

We begin with the assumption that Planet Nine is acting as an exterior perturber in low order mean-motion resonances with one or more detached Kuiper Belt Objects; we wish to derive a probability distribution for Planet Nine's semi-major axis conditional on this assumption. Eleven KBOs with semi-major axes $a > 200$ AU and perihelion distances $q > 30$ AU are considered. The objects are listed in Table~\ref{tab:tab1} with parameters obtained from the Minor Planet Center. 2004 FE\textsubscript{72} has been excluded because its $\sim$ 2155 AU semi-major axis is over four times larger than the average member of this population. We have also excluded uo3L91, as the object's discovery was not yet announced at the time these calculations were performed.

\begin{table}[!h]
\caption{Heliocentric semi-major axes, perihelion distances, inclinations, arguments of perihelion, longitudes of ascending node, and longitudes of perihelion of 11 KBOs with $a > 200 \ \mathrm{AU}$ and $q > 30 \ \mathrm{AU}$} \label{tab:tab1}
\begin{center}
\begin{tabular}{ c | c c c c c c } 
 \hline
 \hline
 Object & $a$ [AU] & $q$ [AU] & $i$ [$^{\circ}$] & $\omega$ [$^{\circ}$] & $\Omega$ [$^{\circ}$] & $\varpi$ [$^{\circ}$] \\
 \hline
 2002 GB\textsubscript{32} & 217.9 & 35.3 & 14.2 & 37 & 177 & 214 \\
 2000 CR\textsubscript{105} & 226.1 & 44.3 & 22.7 & 317.2 & 128.3 & 85.5 \\
 2001 FP\textsubscript{185} & 226.9 & 34.3 & 30.8 & 7 & 179.3 & 186.3 \\
 2012 VP\textsubscript{113} & 260.8 & 80.3 & 24.1 & 292.8 & 90.8 & 23.6 \\
 2014 SR\textsubscript{349} & 289 & 47.6 & 18 & 341.4 & 34.8 & 376.2 \\
 2013 FT\textsubscript{28} & 310.1 & 43.6 & 17.3 & 40.2 & 217.8  & 258 \\
 2004 VN\textsubscript{112} & 317.7 & 47.3 & 25.6 & 327.1 & 66 & 33.1 \\
 2013 RF\textsubscript{98} & 350.0 & 36.1 & 29.6 & 311.8 & 67.6 & 379.4 \\
 2010 GB\textsubscript{174} & 369.7 & 48.8 & 21.5 & 347.8 & 130.6 & 118.4 \\
 2007 TG\textsubscript{422} & 483.5 & 35.6 & 18.6 & 285.7 & 112.9 & 38.6 \\
 Sedna & 499.4 & 76.0 & 11.9 & 311.5 & 144.5 & 96\\
\hline
\end{tabular}
\end{center}
\end{table}

\begin{figure*}[!htbp]
\begin{centering}
\epsscale{1}
\plotone{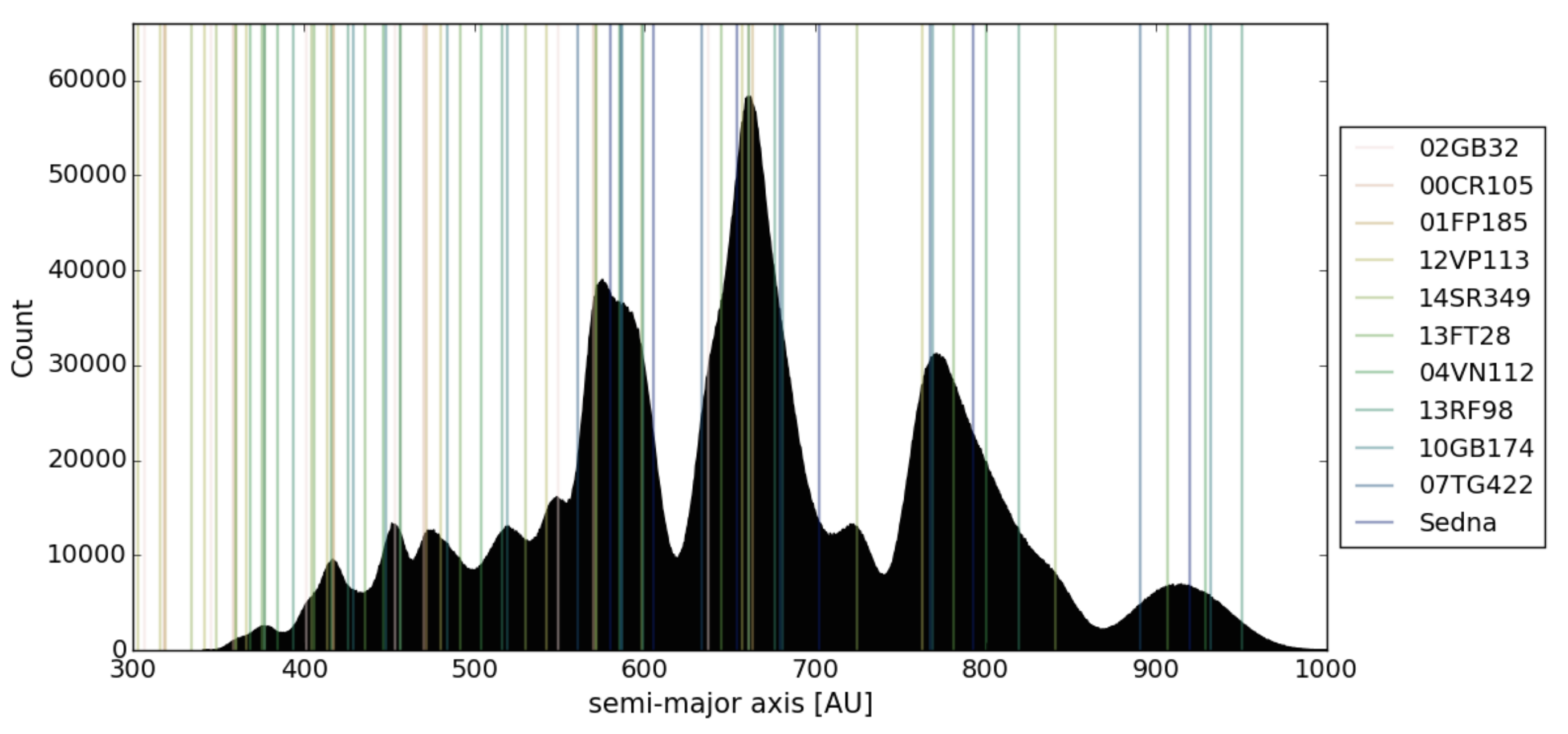}
\caption{A semi-major axis distribution for Planet Nine assuming small integer ratio mean-motion resonances with any number of the 11 KBOs listed in the box to the right of the figure. (The KBOs are sorted by increasing semi-major axes. We use short-hand notation in which ``02GB32'', for example, corresponds to the KBO named 2002 GB\textsubscript{32}.) The plotted distribution of 2 billion samples uses orbital period ratios involving integers up to $N=5$ and a truncated Gaussian mass prior distribution with mean $10 M_{\oplus}$, standard deviation $5 M_{\oplus}$, and bounds $5-20 M_{\oplus}$. There is a noticeably tall and sharp peak at $a\sim$ 660 AU, which, in the range 653-663 AU alone, is associated with MMRs involving five KBOs: Sedna in 3:2, 2000 CR\textsubscript{105} in 5:1, 2012 VP\textsubscript{113} in 4:1, 2004 VN\textsubscript{112} in 3:1, and  2001 FP\textsubscript{185} in 5:1. Increasing the value of the integer $N$ yields more bodies in this narrow semi-major axis range.}
\end{centering}
\label{Figure1}
\end{figure*}

 The semi-major axes of the 11 KBOs yield a corresponding probability distribution for Planet Nine's semi-major axis, if MMRs are to be present. We consider orbital period ratios of integers up to and including the integer parameter $N$. (For example, the MMR ratios associated with $N=4$ are, in order of increasing size, 4/3, 3/2, 2/1, 3/1, and 4/1.)

We place additional restrictions on the semi-major axis by assuming a prior distribution on Planet Nine's mass using the constraints derived by \cite{2016ApJ...824L..23B}. They performed extensive N-body simulations of a synthetic scattered disk of eccentric planetismals, searching over a grid of Planet Nine parameter space for best agreement with the observed spatial clustering. The results favor masses in the range $[5 M_{\oplus}, 20 M_{\oplus}]$, with a most likely value of $m \sim 10 M_{\oplus}$, and a semi-major axis bounded by $ a \sim$ [200 AU + 30$m/M_{\oplus}$, 600 AU + 20$m/M_{\oplus}$]. In the calculation of the semi-major axis probability distribution that follows, we draw on Brown \& Batygin's constraints and use as a mass prior distribution a truncated Gaussian with mean $10 M_{\oplus}$ and standard deviation $\sigma$ (a parameter), and lower and upper bounds of  $5 M_{\oplus}$ and $20 M_{\oplus}$. Varying $\sigma$ allows more or less weight to be placed near the $10 M_{\oplus}$ best estimate.

With this mass prior distribution in hand, we create a physically realistic semi-major axis probability distribution as follows. The maximum MMR integer, $N$, and mass distribution standard deviation, $\sigma$, are parameters of this procedure. 
\begin{enumerate}
\item Draw a mass sample, $m_i$, from the truncated Gaussian prior distribution.
\item Randomly select one of the 11 KBO semi-major axes, $a_{\mathrm{KBO}}$, and one of the allowed MMR integer ratios, $r$. For example, if $N=4$, one ratio $r$ is selected from the set $\left\{4/3, 3/2, 2/1, 3/1, 4/1\right\}$. All resonances are assumed to be equally likely.
\item Calculate the Planet Nine semi-major axis for exact resonance: $a_i = r^{2/3} a_{\mathrm{KBO}}$.
\item Check the Brown \& Batygin semi-major axis bound defined by the mass $m_i$. If $a_i$ lies within {[200 AU + 30$m_i/M_{\oplus}$, 600 AU + 20$m_i/M_{\oplus}$]}, proceed to step 5. If it does not, discard the sample and resume the process from step 1. \item To account for the MMR libration width surrounding exact resonance, draw a Gaussian perturbation to the value of $a_i$. The perturbation is consistent with the inner test particle's libration width as defined by \cite{2016ApJ...824L..22M},\footnote{We note that this $\Delta a_{\rm res}$ measurement is an order-of-magnitude estimate because it was derived in the formalism of the circular restricted three-body problem. The estimate, however, should be sufficient for the purposes of this calculation.}
\begin{equation}
\Delta a_{\rm res} \simeq 0.007\, a_{\rm KBO}\left|\frac{m_i \ a_{\rm KBO} \ {\cal A}}{3 \ M_{\oplus} \ a_i}\right|^{1/2}\, .
\end{equation}
In general, the coefficient $\cal A$ is a complicated function of the KBO semi-major axis, eccentricity, and inclination. It exhibits, however, only a modest dependence on the particular commensurability being considered. We follow Malhotra et al. and adopt the numerical value ${\cal A}=3$. 
\item Repeat steps 1-5 until a semi-major axis distribution of the desired size has been constructed.
\end{enumerate}

\begin{figure*}[!htbp]
\begin{centering}
\epsscale{0.9}
\plotone{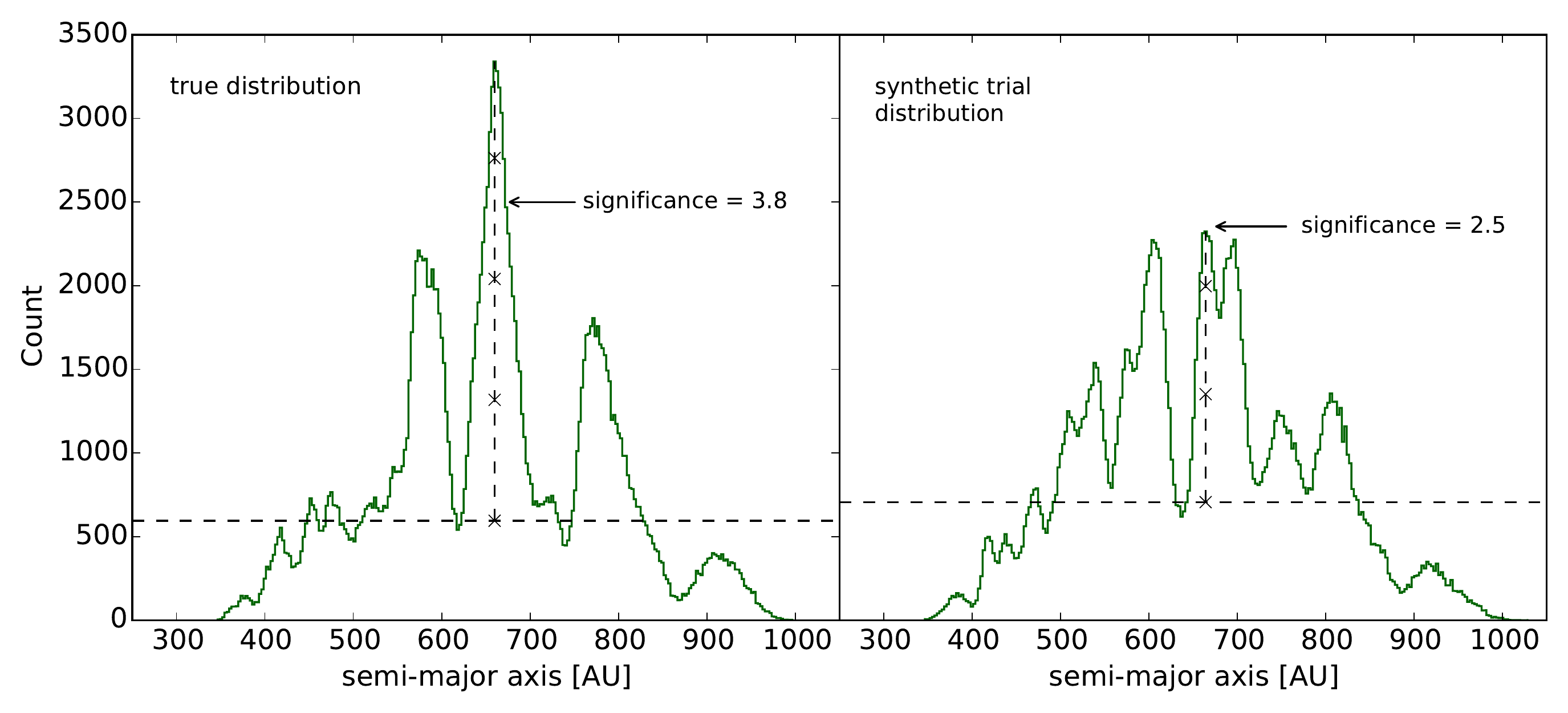}
\caption{Semi-major axis distribution histograms to illustrate the Monte Carlo peak significance simulation. On the left is the true distribution corresponding to the observed KBO semi-major axes. (In effect, this histogram is simply a coarser binning of Figure 1.) On the right is an example Monte Carlo trial distribution corresponding to a set of semi-major axes randomly drawn within the range of the true KBO semi-major axes. Both histograms correspond to $N=5$ and $\sigma = 5$, and they each have 200,000 samples and 250 bins. The horizontal dashed lines are at the medians of the distributions. The ``x'' markers on the vertical dashed line correspond to standard deviations above the median. The labeled significance measurements are thus the number of standard deviations of the highest peak above the median. }
\end{centering}
\label{Semimajor_axis_dist_side_by_side}
\end{figure*}

\begin{figure*}[!htbp]
\begin{centering}
\epsscale{0.9}
\plotone{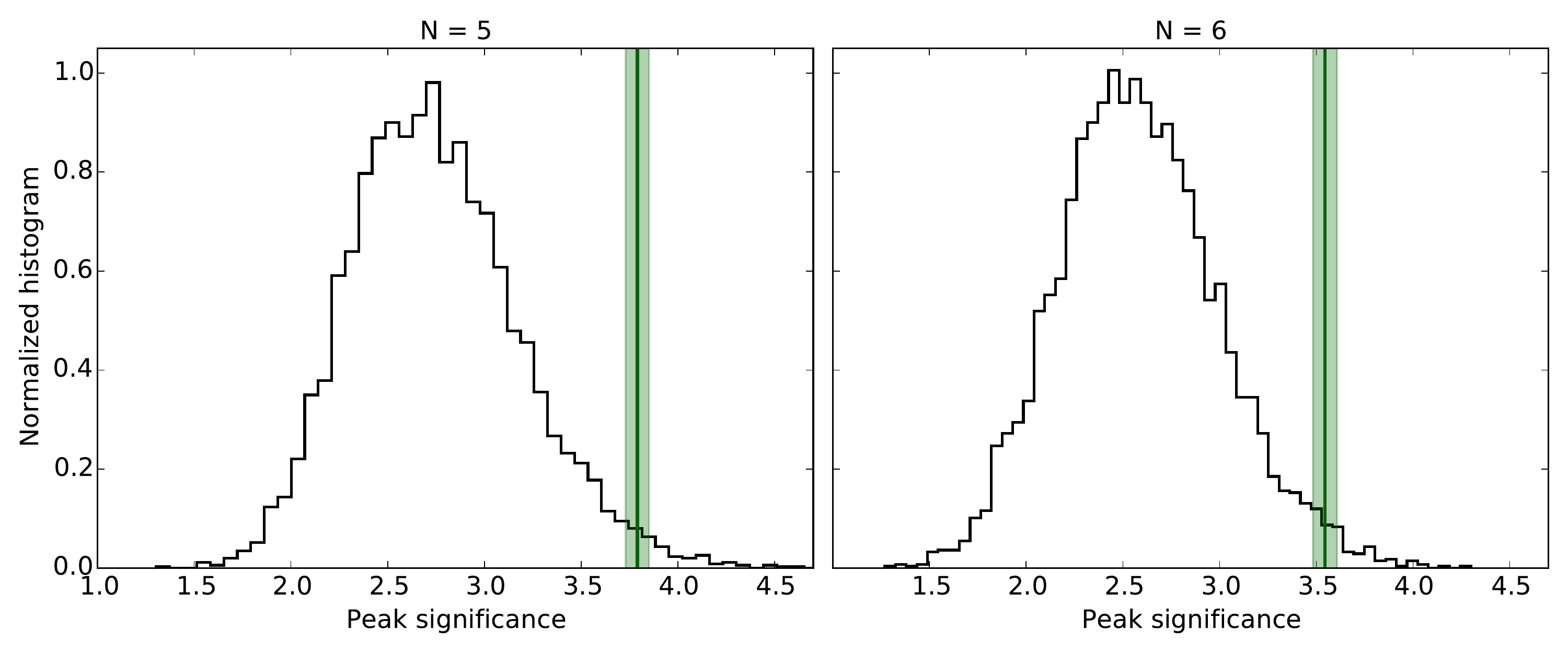}
\caption{Histograms of the significances of the highest peak in 5000 synthetically generated semi-major axis distributions. The cases of $N = 5$ and $N = 6$ are shown with $\sigma = 5$ in each case. The green line indicates the mean significance of the $a \sim 660$ AU peak in the true distribution with the light band giving $1\sigma$ uncertainties. For $N = 5$, 1.6\% of samples lie above the peak significance in the true distribution. For $N = 6$, it is 1.7\%.}
\label{Peak_significance}
\end{centering}
\end{figure*}

In Figure 1, we present the semi-major axis distribution for period ratios involving integers up to $N=5$ and a standard deviation on the mass prior equal to $\sigma = 5 M_{\oplus}$. The figure is a histogram of 2 billion semi-major axis samples.  We denote all values of perfect resonance with vertical colored lines. There is a sharply-defined peak at $a \sim 660$ AU, corresponding to a 3:2 resonance with Sedna, among other small integer resonances. The resonance with Sedna occurs at $a \sim 654$ AU. Because Sedna has the most well-determined semi-major axis, we will use $a \sim 654$ AU to refer to this peak in the remainder of the analysis. The peak is noticeably higher than the others in the distribution, prompting consideration of its statistical significance. That is, given that the real number line is dense with integer ratios, what are the chances of obtaining a peak as tall or taller at random?

We answer this question with a Monte Carlo simulation. In each trial of the simulation, we generate a semi-major axis distribution using the steps outlined above. However, instead of using the true KBO semi-major axes, we use synthetic KBO semi-major axes randomized in the range defined by the semi-major axes of the true KBOs. Each trial uses just one set of randomized semi-major axes. Whereas Figure 1 was generated with 2 billion samples, the trial semi-major axis distributions use only 200,000 samples for computational feasibility. The reduction in sample size has little impact on the calculation we wish to perform. The 200,000 samples of a given trial distribution are organized into 250 bins. An example synthetic trial distribution is shown in the right panel of Figure~\ref{Semimajor_axis_dist_side_by_side}. The left panel shows the true distribution using the actual KBO semi-major axes.

We wish to quantify the percentage of trial distributions that contain a peak as significant as the $\sim$654 AU peak in the true distribution. For each trial distribution, we thus measure the significance of the highest peak as the number of standard deviations of the peak bin height above the median height. A visual depiction of this quantity is shown in Figure~\ref{Semimajor_axis_dist_side_by_side}. Likewise we perform the same peak significance measurements on the true semi-major axis distribution (that corresponding to the observed KBO semi-major axes) and find the mean of 200 measurements. Finally, we calculate the percentage of the synthetic trials that had peak significances greater than the mean peak significance of the true distribution. 

\begin{comment} 

Peak significance results
	normal resonant width			
			max_int	
		    5	6	7
sigma	5	1.6	1.1	1.6
	    7	1.9	1.7	3.1

	half resonant width			
			max_int	
		    5	6	7
sigma	5	2	1.8	1.9
	    7	2.8	2.9	1.7
	    
\end{comment}

Figure~\ref{Peak_significance} displays histograms of the peak significances for 5,000 Monte Carlo trials for the cases of $N=5$ and $N=6$ and $\sigma = 5 M_{\oplus}$ in each case. For $N=5$, only 1.6\% of synthetic samples have peak significances greater than the mean true peak significance, and for $N=6$, it is 1.7\%. To investigate the strength of this result, we repeated our analysis for a collection of other reasonable parameters. Table~\ref{tab:tab2} lists the percentages for $\sigma = 5 M_{\oplus}$ and for a case with less weight near $m = 10M_{\oplus}$, given by $\sigma = 7 M_{\oplus}$. The percentages are quite similar, although in general, the tighter the mass prior distribution, the stronger the result. Finally, to ensure that the result is not an artifact of the assumed resonant width, we decreased $\Delta a_{\rm res}$ to half its size. On average, this results in percentages less than 0.5\% larger.

\begin{table}[!h]
\caption{Percentage of Monte Carlo synthetic trial distributions with peak significances greater than the significance of the true distribution} \label{tab:tab2} 
\begin{center}
\begin{tabular}{ c | c c c } 
 \hline
 \hline
  & $N=5$ & $N=6$ & $N=7$ \\ 
 \hline
 $\sigma = 5 M_{\oplus}$ & 1.6\% & 1.7\% & 1.6\% \\
 $\sigma = 7 M_{\oplus}$ & 1.9\% & 1.7\% & 3.1\% \\
 \hline
\end{tabular}
\end{center}
\end{table}

In summary, when taking into account prior estimates of Planet Nine's mass and mass/semi-major axis relationship, it appears that the KBO semi-major axes exhibit commensurabilities with a Planet Nine semi-major axis of $a\sim654$ AU. We have quantified the low probability that the relation arises by chance. Furthermore, placing a narrower prior on the $10 M_{\oplus}$ estimate most favored by Brown \& Batygin only strengthens the case. We therefore devote the rest of our analysis to exploring the plausibility of Planet Nine residing in an $a \sim 654$ AU orbit. We stress that our results rest on this fundamental assumption and that they lose their import if the peak in the semi-major axis distribution has arisen by chance.

\section{Numerical Methods}

If Planet Nine has $a \sim 654$ AU and participates in mean-motion resonances with a subset of the distant KBOs, the MMRs necessarily impose constraints on both its orbital parameters and its current location. For example, given that Sedna is currently near its perihelion approach, a 3:2 resonant configuration would imply that Planet Nine's mean anomaly is close to one of three locations: $M=60^\circ$, $M=180^\circ$, or $M=300^\circ$. Moreover, direct numerical calculations can be used to explore the consequences of the requirement that resonances exist, thereby increasing the constraints on all of Planet Nine's orbital parameters.

Most orbital configurations that are consistent with Planet Nine's nominal parameters lead to rapid orbital evolution of the most distant KBOs. For example, \citet{2016MNRAS.460L.123D} show that when the baseline Planet Nine model of \citet{2016AJ....151...22B} is adopted, the orbits of 2004 VN\textsubscript{112}, 2007 TG\textsubscript{422}, and 2013 RF\textsubscript{98} are destabilized within a few tens of Myr. Furthermore, the apsidal alignment is not maintained for arbitrary instances that are consistent with the nominal model. However, \cite{2016MNRAS.460L.123D} show that slight modifications that enforce $\Delta\varpi = 180^{\circ}$ in the \cite{2016AJ....151...22B} model lead to improved KBO stability. If Planet Nine has $a\sim 654$ AU, orbital parameter domains must exist where the KBOs are relatively stable and where apsidal alignment is maintained for long periods. 

We have conducted a comprehensive parameter search by numerically integrating a set of KBOs under the influence of Planet Nine and seeking out domains of orbital parameter space where KBO large-scale stability and apsidal alignment are maintained. In our integrations we included the eleven KBOs with semi-major axes $a > 200$ AU and perihelion distances $q > 30$ AU listed in Table~\ref{tab:tab1}. We restricted, however, the constraint on apsidal alignment to only seven of these with $a>250$ AU. The only member with $a>250$ AU that we exclude is 2013 FT\textsubscript{28}, which is anti-aligned from the rest of the KBOs \citep{2016AJ....152..221S}. Provided the results of the KBO orbital evolution under the influence of the distant planet, we wish to observe these seven KBOs cluster in their arguments and longitudes of perihelion. We first define simple merit functions that allow us to parameterize the degree to which clustering is maintained in a time-averaged sense.

\begin{comment}

If $\left |\Delta a_i/a_{i0} \right |$ is the maximum absolute fractional deviation of the $i^{th}$ KBO's semi-major axis, $x = \frac{1}{N} \sum_{i=1}^{19}\left |\Delta a_i/a_{i0} \right |$ is a measure of the average semi-major axis instability over the nineteen KBOs we are tracking. We want to permit orbital motion to some degree with little penalty; we thus define the orbital stability merit function as the logistic function $f_1(x; k, x_0) = 1/(1 + e^{k(x-x_0)})$. We found $x_0 = 0.25$ and $k=30$ to be suitable parameters. As shown in Figure 2, the merit is close to 1 for average maximum absolute fractional deviations $x < 0.25$ and drops off sharply thereafter. 

\end{comment}

\subsection{Quantifying apsidal confinment}\label{clustering_metric}

A useful quantification of clustering stems from the degree to which the KBO longitudes of perihelion, $\varpi_i$, are spread about anti-alignment with Planet Nine.  If 
\begin{equation}
\left | \varpi_i - \varpi_{P9} - 180^\circ \right | = \left | \Delta \varpi_i - 180^\circ \right |
\end{equation}
is the absolute deviation of the $i^{th}$ KBO's longitude of perihelion from perfect \textit{anti-alignment} with Planet Nine (in a range from $0^{\circ}-180^{\circ}$), then 
\begin{equation}
m= \underset{1 \leq i \leq 7}{\rm median}{\left | \Delta \varpi_i - 180^\circ \right |}
\end{equation}
is a measure of the typical deviation of the seven KBOs at one slice in time. The median over T time slices, 
\begin{equation}
s = \underset{1 \leq t \leq T}{\rm median} \left \{ m \right \}\, ,
\end{equation}
is thus an average measure of the scatter across the duration of the integration. We wish to favor configurations that produce small values of $s$. A suitable merit function is  
\begin{equation}
f(s) = e^{-s^2/1000}\, ,
\end{equation}
where the constant 1,000 was chosen through experimentation.

\begin{table*}[!t]
\caption{Initial parameter distributions} \label{tab:tab3} 
\begin{center}
\begin{tabular}{ c c } 
 \hline
 \hline
 Parameter & Distribution \\
 \hline
 semi-major axis, $a$ $[\textrm{AU}]$ & $a \sim \mathcal{N}(654, 20, 590, 720)$ \\ 
 mass, $m$ $[M_{\oplus}]$ & $m \sim \mathcal{N}[10, 7.5, (a - 600)/20,$ $(a - 200)/30]$ \\
 eccentricity, $e$ & $e \sim \mathcal{N}[0.5 - 0.05(m - 10), 0.2, 0.1, 0.9]$ \\ 
 inclination, $i$ $[^{\circ}]$ & $i \sim \mathcal{N}(30, 20, -10, 70)$ \\
 longitude of ascending node, $\Omega$ $[^{\circ}]$ & $\Omega \sim \mathcal{U}[0, 360]$ \\
 argument of perihelion, $\omega$ $[^{\circ}]$ & $\omega \sim \mathcal{N}(230 - \Omega, 25)$ \\
 mean anomaly, $M$ $[^{\circ}]$ & $M \sim \mathcal{U}[0, 360]$\\
 \hline
\end{tabular}
\end{center}
\end{table*}

A conservative merit function only demands KBO clustering rather than both clustering and orbital anti-alignment with Planet Nine. This may be parameterized by alignment of the KBO argument of perihelion vectors,
\begin{equation}
\vec{e_{\omega}} = (\cos\omega)\hat{i} + (\sin\omega)\hat{j}\, .
\end{equation}
Provided a set of orbital evolutions for seven KBOs, we quantify the clustering as follows. We first calculate the seven bodies' unit argument of perihelion vectors $\vec{e_1}(t), ..., \vec{e_7}(t)$ and their mean unit vector $\vec{e_m}(t)$ at all times $t$. We compute the sum of the time-varying dot products, 
\begin{equation}
S_{\omega}(t) = \sum_{i=1}^{7}\vec{e_i}(t) \cdot \vec{e_m}(t)\, ,
\end{equation}
which quantifies the degree to which the vectors are clustered about their mean. If a KBO gets ejected from the system before the integration is over, we set $\vec{e_i}(t)$ to zero, such that this KBO stops contributing to $S_{\omega}(t)$. We also evaluate the time-varying anti-alignment with respect to Planet Nine as
\begin{equation}
A_{\omega}(t) = \vec{e_m}(t) \cdot \vec{e_{P9}}(t)\,. 
\end{equation}
If the seven bodies are perfectly anti-aligned with respect to Planet Nine for all times, then $S_{\omega}(t)=7$ and $A_{\omega}(t)=-1$. The present-day value of $S_{\omega}$ for the seven KBOs we are considering is $S_{\omega}(0) = 6.52$. The time average $\left<{S_{\omega}(t)}\right>$ is an effective measure of confinement. Assuming that the present-day observed configuration of the KBOs is not a unique moment in time, $\left<{S_{\omega}(t)}\right>$ should be close to $S_{\omega}(0)$. Furthermore, an integration that maintains KBO anti-alignment with P9 will yield $\left<{A_{\omega}(t)}\right> \approx -1$. 

Analogous metrics may be constructed with the longitude of perihelion vectors,
\begin{equation}
\vec{e_{\varpi}} = (\cos\varpi)\hat{i} + (\sin\varpi)\hat{j}\, ,
\end{equation}
for which the present-day value is $S_{\varpi}(0) = 5.63$. A successful configuration must produce clustering in both $\omega$ and $\varpi$ in order to match the observations. We note that it is insufficient to enforce clustering in $\varpi$ or the Runge-Lenz eccentricity vectors alone, as these metrics do not guarantee confinement in $\omega$.

In the analysis that follows, the merit function $f(s)$ will be used to guide a Markov Chain Monte Carlo procedure; $\left<{S_{\omega}(t)}\right>$, $\left<{A_{\omega}(t)}\right>$, $\left<{S_{\varpi}(t)}\right>$, and $\left<{A_{\varpi}(t)}\right>$ will be used to further analyze the resulting samples.

\subsection{DE-MCMC parameter space exploration}\label{DEMCMC}

Having defined merit functions by which we can judge the KBO orbital evolution associated with a given set of Planet Nine parameters, we next discuss how these parameters are sampled. We begin with initial parameter distributions that are motivated by the strong suggestion of commensurability in Section~\ref{section2} and by the constraints provided by the simulations described by \cite{2016ApJ...824L..23B}. The distributions are listed in Table~\ref{tab:tab3}, where we define $\mathcal{N}(\mu, \sigma, a, b)$ as a truncated Gaussian probability distribution with mean $\mu$, standard deviation $\sigma$, and lower and upper bounds, $a$ and $b$. $\mathcal{N}(\mu, \sigma)$ and $\mathcal{U}[a,b]$ are Gaussian and uniform distributions, respectively. 

The dependence of the $m$ distribution on $a$ corresponds to the constraints derived by \cite{2016ApJ...824L..23B}. This mass/semi-major axis relationship was also employed in Section~\ref{section2} when we used mass constraints within the calculation of the MMR semi-major axis distribution. Similarly, the relationship of the eccentricity distribution to the mass has been adapted from Brown \& Batygin's constraints. The main feature is the negative correlation between $e$ and $m$. Finally, we note that the longitude of perihelion has been constrained to $\varpi \sim \mathcal{N}(230^{\circ}, 25^{\circ})$ in accord with anti-alignment with the KBOs. $\omega$ and $\Omega$ are allowed to vary through a full $360^{\circ}$, subject to the constraint that $\varpi = \omega + \Omega$.

\begin{comment}
These were for the 200 Myr integrations:
$a \sim \mathcal{N}(660\textrm{ AU}, 20\textrm{ AU}, 590\textrm{ AU}, 720\textrm{ AU})$, \\
$m/M_{\oplus} \sim \mathcal{N}(10, 7.5, [a - 600\textrm{ AU}]/20\textrm{ AU},$
$[a - 200\textrm{ AU}]/30\textrm{ AU})$, \\
$e \sim 0.5 - 0.05(m - 10) + \mathcal{N}(0, 0.2)$, \\
$i \sim \mathcal{N}(30^{\circ}, 15^{\circ}, -5^{\circ}, 65^{\circ})$, \\
$\omega \sim \mathcal{N}(140^{\circ}, 20^{\circ}, 65^{\circ}, 215^{\circ})$, \\
$\Omega \sim \mathcal{N}(113^{\circ}, 20^{\circ}, 40^{\circ}, 190^{\circ})$, \\
$M \sim \mathcal{U}[0^{\circ}, 360^{\circ}]$.
\end{comment}

Using these initial parameter distributions to seed a collection of Markov Chains, we carried out a parameter exploration using a Differential Evolution Markov Chain Monte Carlo (DE-MCMC) algorithm \citep{TerBraak2006}. The DE-MCMC technique is most often used to sample from a posterior parameter distribution, typically for Bayesian inference; it iterates many Markov Chains in parallel and uses the inter-chain differences to inform the parameter jumps from one iteration to the next. In our analysis, however, the merit function takes the place of a likelihood function. As a result, this technique will \textit{not} produce true posterior parameter distributions, as the merit function is not a true probability density function. Rather, the DE-MCMC is simply a tool that we use to efficiently sample the 7-dimensional parameter space, assuming large parameter covariance and non-smooth parameter space structure, both of which DE-MCMC is helpful in accounting for. 

We ran in parallel a large collection of DE-MCMC instances of 5 chains. We used jumping parameters $\gamma = 0.1$ and $\gamma_s =  0.001$ \citep{TerBraak2006}, such that one chain's proposal jump during a given iteration was approximately one-tenth of the vector between two other randomly selected chains. As stated above, the merit function $f(s)$ took the place of the likelihood. The sampling of $e$ and $\omega$ used the transformations $e\cos(\omega)$ and $e\sin(\omega)$; the other orbital elements were unaltered. 

For all of our N-body dynamical integrations, we used the hybrid symplectic-Bulirsch-Stoer algorithm from the Mercury N-body integration package \citep{1999MNRAS.304..793C}. We modeled the gravitational potential of the inner three giant planets (Jupiter, Saturn, and Uranus) as an orbit-averaged quadrupole by giving the Sun a physical radius equal to Uranus's orbital radius and a $J_2$ moment equal to \citep{Burns1976}
\begin{equation}
J_2 = \frac{{m_J {a_J}^2}+{m_S {a_S}^2}+{m_U {a_U}^2}}{2 M_{\odot} {a_U}^2}\, .
\end{equation}
\vskip 0.15in

Neptune was followed in direct N-body fashion, and the integration timestep was set to one tenth of Neptune's orbital period. The integrations were carried out for one billion years. While this is too short to ensure apsidal confinement lasting over the age of the solar system, it is certainly long enough to observe KBO orbital instability and loss of apsidal confinement due to differential precession (that is, within a configuration in which Planet Nine's orbital parameters are unsuitable). \\

\begin{figure*}[!htbp]
\begin{centering}
\epsscale{1}
\plotone{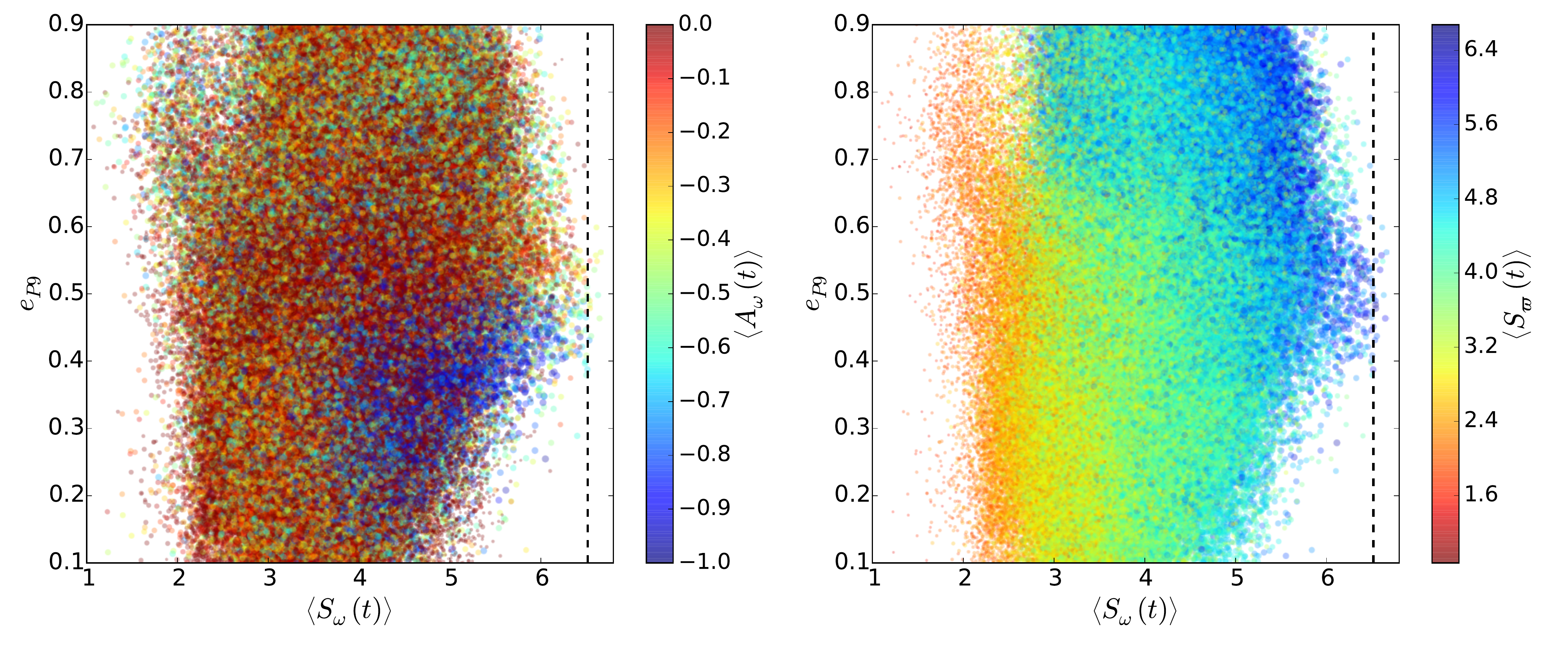}
\caption{The eccentricity of Planet Nine versus $\left<{S_{\omega}(t)}\right>$, a time-averaged measure of clustering in the KBO arguments of perihelion. The dashed vertical line represents the present-day value of the clustering in the seven KBOs, $S_{\omega}(0) = 6.52$. In the left panel, the colorbar $\left<{A_{\omega}(t)}\right>$ quantifies anti-alignment of the arguments of perihelion with Planet Nine, with constant anti-alignment corresponding to -1. ($\left<{A_{\omega}(t)}\right>$ extends to values larger than 0, but the colorbar has been truncated for greater color contrast.) In the right panel, the colorbar $\left<{S_{\varpi}(t)}\right>$ quantifies clustering in the longitude of perihelion vectors, with a present-day value of $S_{\varpi}(0) = 5.63$. The sizes of the points are scaled with the colorbars in order to enhance the regions of anti-alignment and longitude of perihelion clustering. }
\label{ecc_anti_align_and_long}
\end{centering}
\end{figure*}

\subsection{KBO cloning and multiplexing}

While monitoring the orbital clustering of the seven KBOs is straightforward, the KBOs' own orbital uncertainties must also be accounted for. Inspection of the bodies listed in Table~\ref{tab:tab1} using the the JPL Small-Body Database Browser\footnote{\url{http://ssd.jpl.nasa.gov/sbdb.cgi\#top}} indicates that the semi-major axis uncertainties for the distant KBOs are of similar magnitude to the resonant libration widths. If mean-motion resonances are indeed the primary mechanism for the orbital clustering, the uncertainties must be considered when evaluating whether the bodies can coexist in resonance.

\begin{comment}
In section 3.1, we observed that Sedna, 2012 VP\textsubscript{113}, and 2004 VN\textsubscript{112} were capable of libration in their respective 3:2, 4:1, and 3:1 mean-motion resonances, but that the values of $a_{P9}$ at the cores of the resonances varied by a few AU. However, overlapping resonances are easily allowable given the uncertainties in the KBO semi-major axes. For a KBO in a $p+q:p$ MMR, a shift $\Delta a$ in the KBO's semi-major axis yields a corresponding shift $\Delta a_{P9}$ for exact resonance: $\Delta a_{P9} = \left(\frac{p+q}{p}\right)^{2/3} \Delta a$. For example, a shift of $\Delta a \sim$ 1 AU in 2012 VP\textsubscript{113}'s semi-major axis, well within its $\sim 3.3$ AU $1\sigma$ uncertainty, results in $\Delta a_{P9} \sim$ 2.5 AU. This is more than enough to make up for the difference observed in Figure x.
\end{comment}

To accommodate the uncertainties, all of our integrations were performed not only on the KBOs themselves, but also on a set of ``clones''. For each integration, we randomly generated five clones per KBO with identical values for $e_i$, $i_i$, $\Omega_i$, $\omega_i$, and $M_i$ but with a Gaussian perturbation to the semi-major axis, $a_i$, consistent with the currently estimated semi-major axis uncertainties. Ideally, one would draw perturbations on all the KBO orbital elements, but the semi-major axis is clearly the principal parameter in determining the proximity to MMR.\footnote{After generating the data set presented in this paper, we ran separate calculations in which the clones' full set of orbital elements were sampled from the KBO uncertainty covariance matrices. We generated a sample size that is one third as large as the data set presented here. We found no significant differences between the results that only varied the clone semi-major axes and those that sampled all orbital elements.} We retained one copy of the original KBO with no perturbation, for a total of six clones per KBO, or 42 bodies total to be followed in each integration.

\begin{figure*}[!htbp]
\begin{centering}
\epsscale{1.1}
\plotone{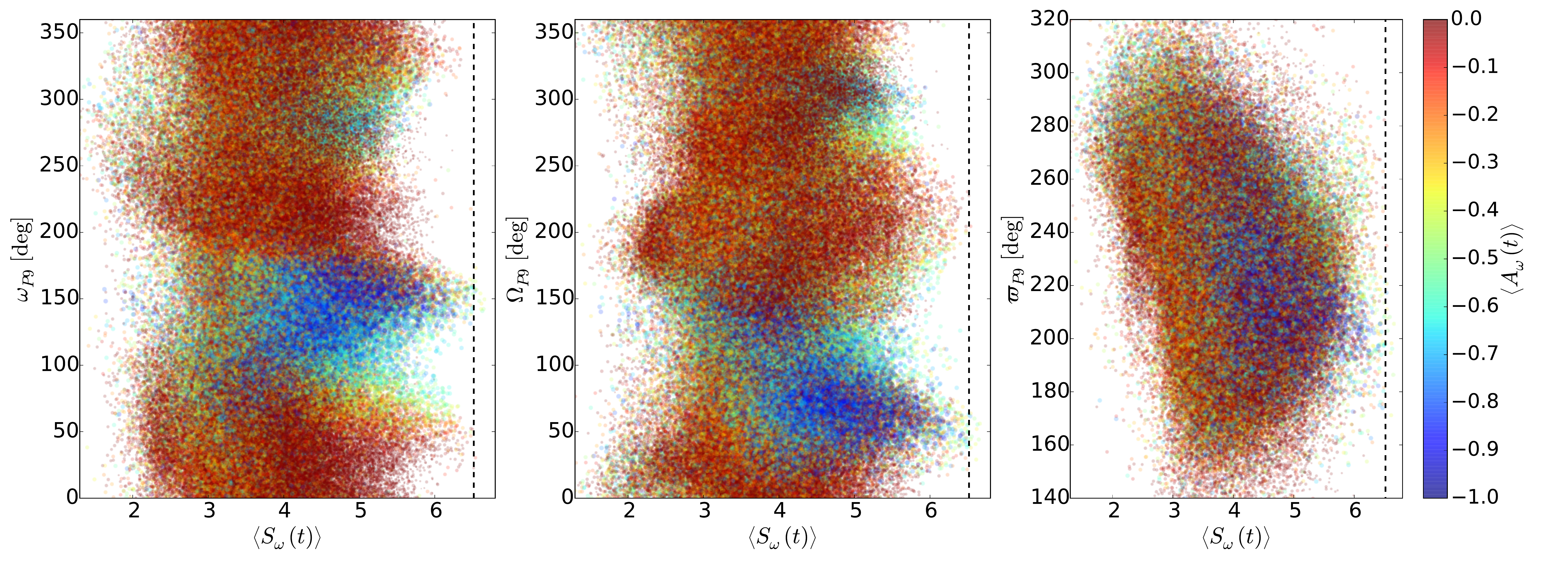}
\caption{The distribution of arguments of perihelion, longitudes of ascending node, and longitudes of perihelion. See the Figure~\ref{ecc_anti_align_and_long} caption for details on the plotting parameters.}
\label{omega_Omega_varpi_anti_align}
\end{centering}
\end{figure*}

A benefit of representing seven KBOs as 42 independent bodies is a dramatic increase in sample size enabled through a multiplexing process. In keeping with the hypothesis of \cite{2016ApJ...824L..22M}, we expect that the clone with semi-major axis closest to  exact resonance will generally perform the best at maintaining apsidal alignment in time. For a given integration, we seek the set of seven best-performing clones. There are $6^7 = 279,936$ such sets of possible combinations. In effect, each dynamical integration of 42 bodies that is performed can be recast as a set of 279,936 independent integrations of seven bodies, each of them with the same set of parameters for Planet Nine but with different parameters for the seven KBOs of interest.

To summarize, we \textit{a posteriori} generated the 279,936 multiplexed samples from each billion-year integration that was performed using the procedure discussed in Section \ref{DEMCMC}. In each multiplexed sample's combination of seven KBOs, we calculated the measures of argument and longitude of perihelion vector clustering and anti-alignment that were defined in Section \ref{clustering_metric}.

\section{RESULTS}

\subsection{Orbital element constraints}\label{orb_elmt_constraints}

We now present the results of $\sim 250,000$ individual billion-year orbital integrations. Rather than considering all $6^7$ multiplexed samples for each integration, which would produce a prohibitively large set of 70 billion one-Gyr integrations to analyze, we first consider a subset of the best performing samples. This subset consists of each integration's single multiplexed sample that exhibits the largest value of $\left<{S_{\omega}(t)}\right>$, defined in Section~\ref{clustering_metric} as the time average of the sum of the dot products of the seven KBO argument of perihelion vectors with their mean vector. In this procedure, we opted to maximize $\left<{S_{\omega}(t)}\right>$ as opposed to $\left<{S_{\varpi}(t)}\right>$ because, as we will show, obtaining clustering close to the present-day observations is much more challenging for $\omega$ than $\varpi$.

\begin{figure}[!h]
\begin{centering}
\epsscale{1.3}
\plotone{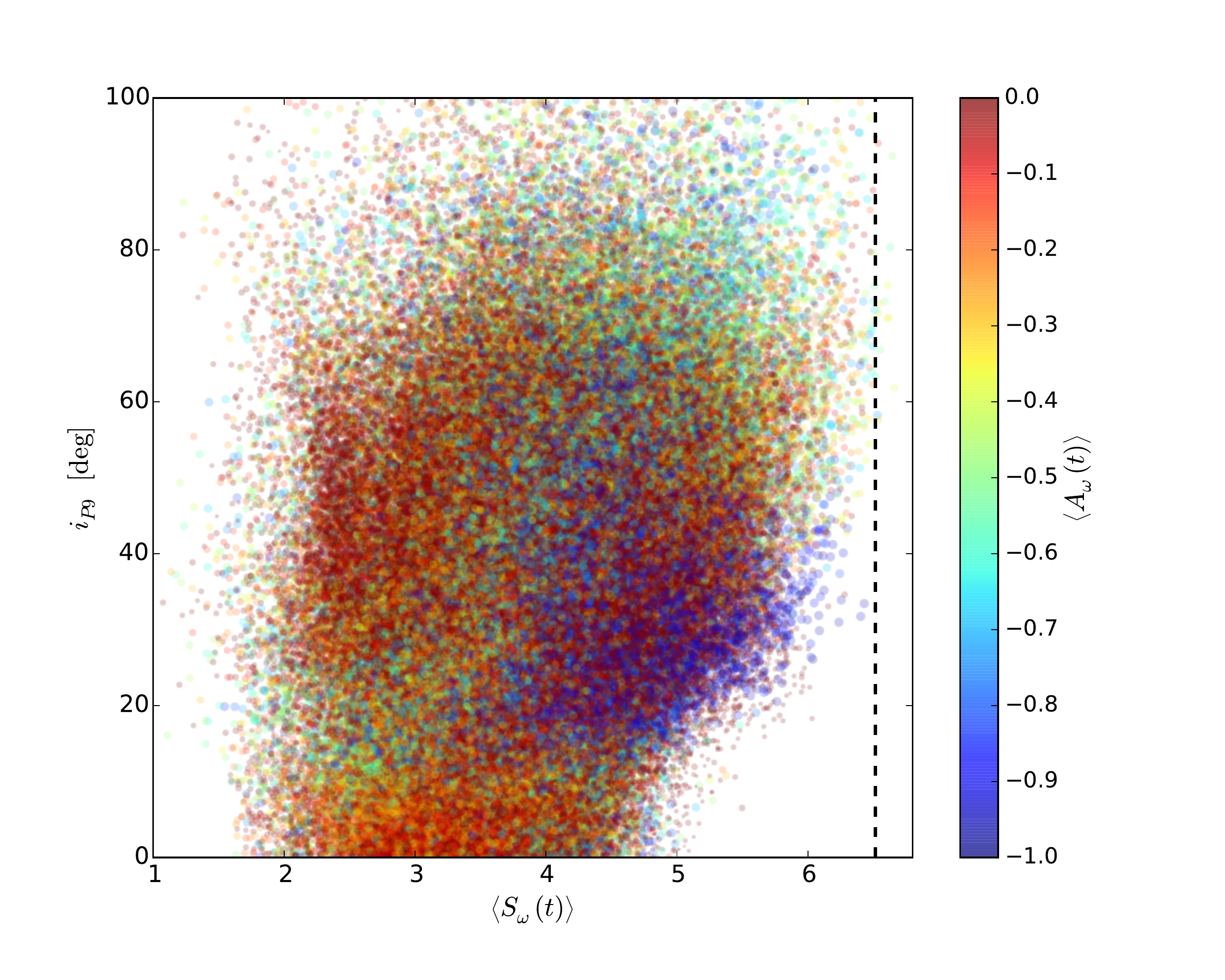}
\caption{The distribution of inclinations. See the Figure~\ref{ecc_anti_align_and_long} caption for details on the plotting parameters.}
\label{inc_anti_align}
\end{centering}
\end{figure}

\begin{figure}[!h]
\begin{centering}
\epsscale{1.3}
\plotone{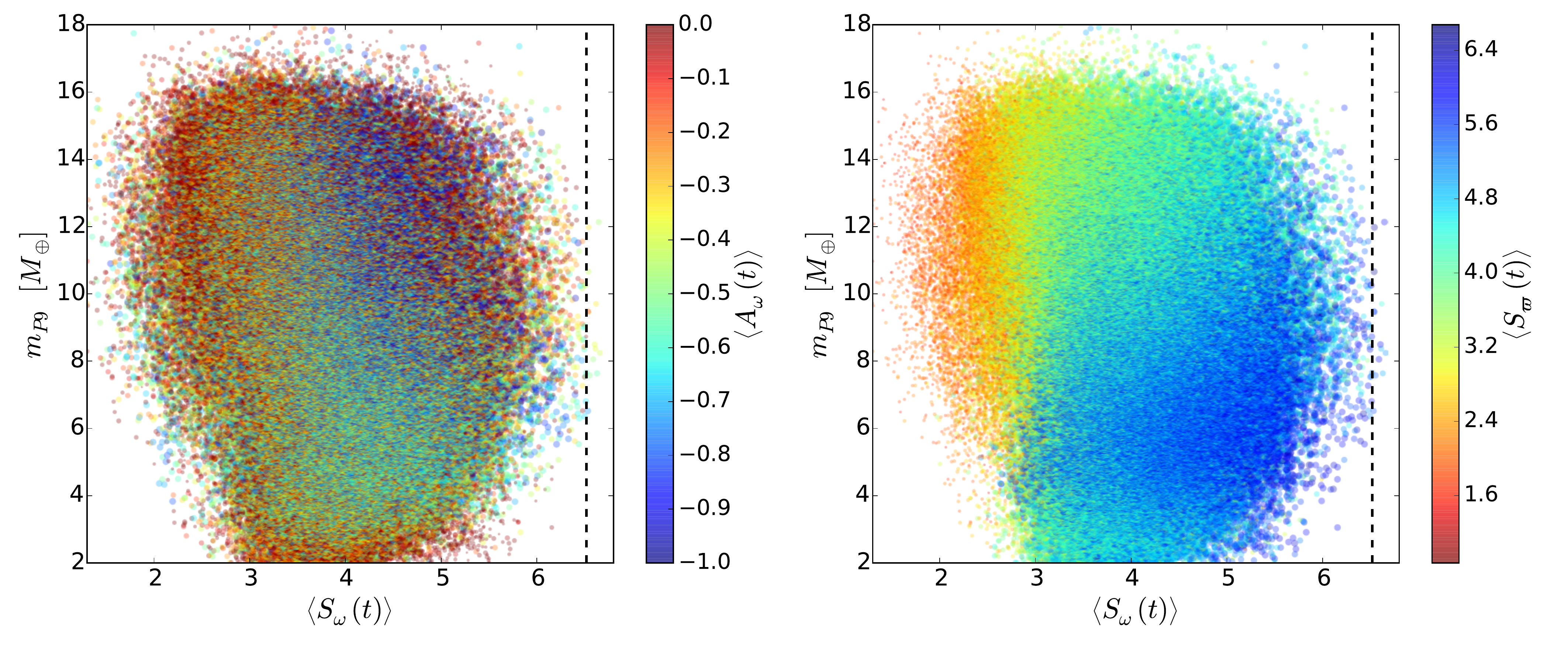}
\caption{The distribution of masses for Planet Nine. See the Figure~\ref{ecc_anti_align_and_long} caption for details on the plotting parameters.}
\label{mass_anti_align_and_long}
\end{centering}
\end{figure}

Figures~\ref{ecc_anti_align_and_long} - \ref{mass_anti_align_and_long} show the performance of the KBOs in maintaining clustering in their arguments of perihelion as a function of Planet Nine's orbital elements. In each scatter plot, the $x$-axis is the argument of perihelion clustering measure $\left<{S_{\omega}(t)}\right>$, with the dashed line marking the present-day value, $S_{\omega}(0) = 6.52$. The colorbar variable $\left<{A_{\omega}(t)}\right>$, as defined in Section~\ref{clustering_metric}, quantifies the degree to which the KBOs maintained anti-alignment with Planet Nine, with $\left<{A_{\omega}(t)}\right>=-1$ being perfect anti-alignment. Figures~\ref{ecc_anti_align_and_long} and ~\ref{mass_anti_align_and_long} include an additional scatter plot using $\left<{S_{\varpi}(t)}\right>$, the longitude of perihelion clustering measurement,  as the colorbar.
%To enhance the regions of anti-alignment, we have scaled the size of the points with $\left<{A_{\omega}(t)}\right>$. 

We first make some general observations. The figures indicate that an optimization scheme of this nature is a viable technique for constraining Planet Nine's orbital elements. Obtaining large values of $\left<{S_{\omega}(t)}\right>$ is challenging, and the vast majority of orbital configurations do not work out, yet there are constrained regions of the parameter space where significant clustering can be maintained.

$\left<{S_{\omega}(t)}\right>$ is generally anti-correlated with $\left<{A_{\omega}(t)}\right>$, although the samples with the largest values of $\left<{S_{\omega}(t)}\right>$ are not exclusively anti-aligned with Planet Nine. We note that anti-alignment with Planet Nine is not required by the KBO observations, which simply exhibit an observed clustering in $\omega$ and $\varpi$. It is difficult to conceive, however, of a configuration that is \textit{not} anti-aligned and that can survive for billions of years. Either secular or resonant confinement mechanisms require $\Delta \varpi \approx 180^{\circ}$, with $\Delta \omega \neq 180^{\circ}$ implying a nodal asymmetry. When longer time integrations are performed, we expect there will be a decrease in the number of samples that perform well without an anti-aligned configuration.

Figure~\ref{ecc_anti_align_and_long} presents the results for the eccentricity. Samples with the largest values of $\left<{S_{\omega}(t)}\right>$ have $e \sim 0.4-0.55$, with the left panel showing that anti-aligned configurations have $e \sim 0.4-0.5$. 
This is slightly lower but not in disagreement with the eccentricities favored by \cite{2016ApJ...824L..23B} for $a \sim 600-700$ AU. The right panel of Figure~\ref{ecc_anti_align_and_long} depicts similar results, except the colorbar now shows clustering in the KBO longitudes of perihelion. Given that $S_{\varpi}(0) = 5.63$, clustering is evidently much easier for $\varpi$ than for $\omega$. Interestingly, a large degree of clustering in $\varpi$ is most easily attainable for highly eccentric orbits.

Figure~\ref{omega_Omega_varpi_anti_align} presents the distributions of $\omega$, $\Omega$, and $\varpi$. As discussed in Section~\ref{DEMCMC}, the initial distribution restricted $\varpi$ to initial anti-alignment with the KBOs but let $\omega$ and $\Omega$ vary through $0^{\circ}-360^{\circ}$. It is unsurprising to see strongest anti-alignment occurring for $\omega \sim 150^{\circ}$, as this corresponds to present-day anti-alignment with the KBOs. It is still significant, however, as in the absence of the perturbing planet, the KBOs randomize their alignment in tens of millions of years due to differential precession. Moreover, the region where $\omega \sim 150^{\circ}$ also produces the strongest clustering in the KBO arguments of perihelion. The longitude of the ascending node displays strongest clustering and anti-alignment for $\varpi \sim 205^{\circ}$. By extension, the node is best at $\Omega \sim 50^{\circ}$.

\begin{comment}
\begin{figure*}[!tbp]
\begin{centering}
\epsscale{1}
\plotone{mass_anti_align_and_long.pdf}
\caption{The distribution of masses for Planet Nine. See the Figure~\ref{ecc_anti_align_and_long} caption for details on the plotting parameters.}
\label{mass_anti_align_and_long}
\end{centering}
\end{figure*}
\end{comment}

The inclination distribution in Figure~\ref{inc_anti_align} illustrates that low inclinations $i \lesssim 20^{\circ}$ are consistently disfavored. Strongest anti-alignment occurs for $30^{\circ} \lesssim i \lesssim 40^{\circ}$, in agreement with the \cite{2016ApJ...824L..23B} estimate, whereas clustering is strongest for very high inclinations. To understand this apparent preference for high inclination configurations, we examined the KBO trajectories of representative samples of this population. The KBOs in these configurations typically undergo large eccentricity oscillations and a rapid increase to highly inclined or retrograde orbits. In general, the bodies are only able to maintain inclinations close to their present-day values for $i \lesssim 40^{\circ}$, although some configurations with $i < 40^{\circ}$ can still cause the KBOs to evolve to high inclinations. This evolution to high inclination and retrograde orbits was discussed in \cite{2016arXiv161004251S} and in \cite{2016ApJ...833L...3B}, 
who showed that Planet Nine is capable of producing the observed population of highly inclined, $a < 100$ AU trans-Neptunian objects through a process involving Kozai-Lidov cycles and close encounters with Neptune.  We also found a moderate correlation between KBO evolution to highly inclined orbits and \textit{nodal} clustering, possibly providing a connection to the observation of nodal clustering in the population of $a < 100$ AU TNOs reported in \cite{2016ApJ...827L..24C}. An extended exploration of these inclination dynamics is saved for a future discussion.

In contrast to some of the other orbital elements, the mass distribution in Figure~\ref{mass_anti_align_and_long} exhibits less constraint. The KBO clustering and anti-alignment are possible for masses throughout the range $6 M_{\oplus} \lesssim m \lesssim 12 M_{\oplus}$. In agreement with previous studies, masses less than $\sim 6 M_{\oplus}$ are strongly disfavored.

The semi-major axis distribution does not display much structure and is not shown here. The KBO semi-major axes are constantly perturbed, however, through the cloning procedure, meaning that the semi-major axis ratio $a_{P9}/a_{KBO}$ or period ratio $P_{P9}/P_{KBO}$ are more physically informative than $a_{P9}$. We pursue an exploration of the period ratio in Section ~\ref{resonance}.

\subsection{Resonant KBOs}\label{resonance}

\begin{figure}[!h]
\begin{centering}
\epsscale{1.25}
\plotone{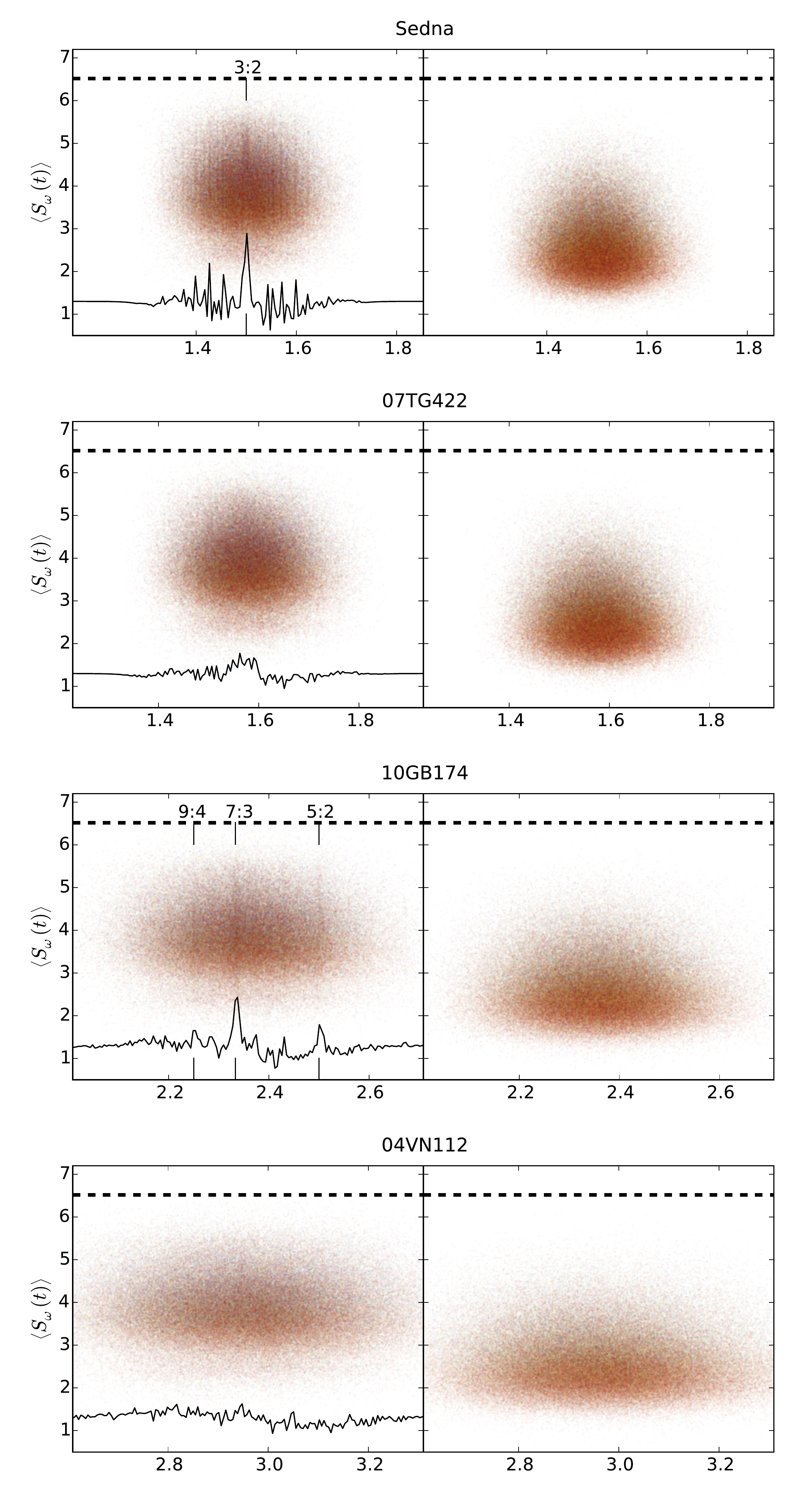}
\caption{The KBO clustering merit  $\left<{S_{\omega}(t)}\right>$ versus the period ratio $P_{P9}/P_{KBO}$ at $t=0$, with coloration indicating the measure of anti-alignment $\left<{A_{\omega}(t)}\right>$ as in Figures~\ref{ecc_anti_align_and_long} -~\ref{mass_anti_align_and_long}. The horizontal dashed lines are at the present-day value, $S_{\omega}(0) = 6.52$. The left panels for each KBO correspond to the best-performing multiplexed samples and the right panels to random multiplexed samples. The black curves in the left panels are differenced histograms as described in the text. This is four of the seven KBOs.}
\label{period_ratio1}
\end{centering}
\end{figure}

\begin{figure}[!t]
\begin{centering}
\epsscale{1.25}
\plotone{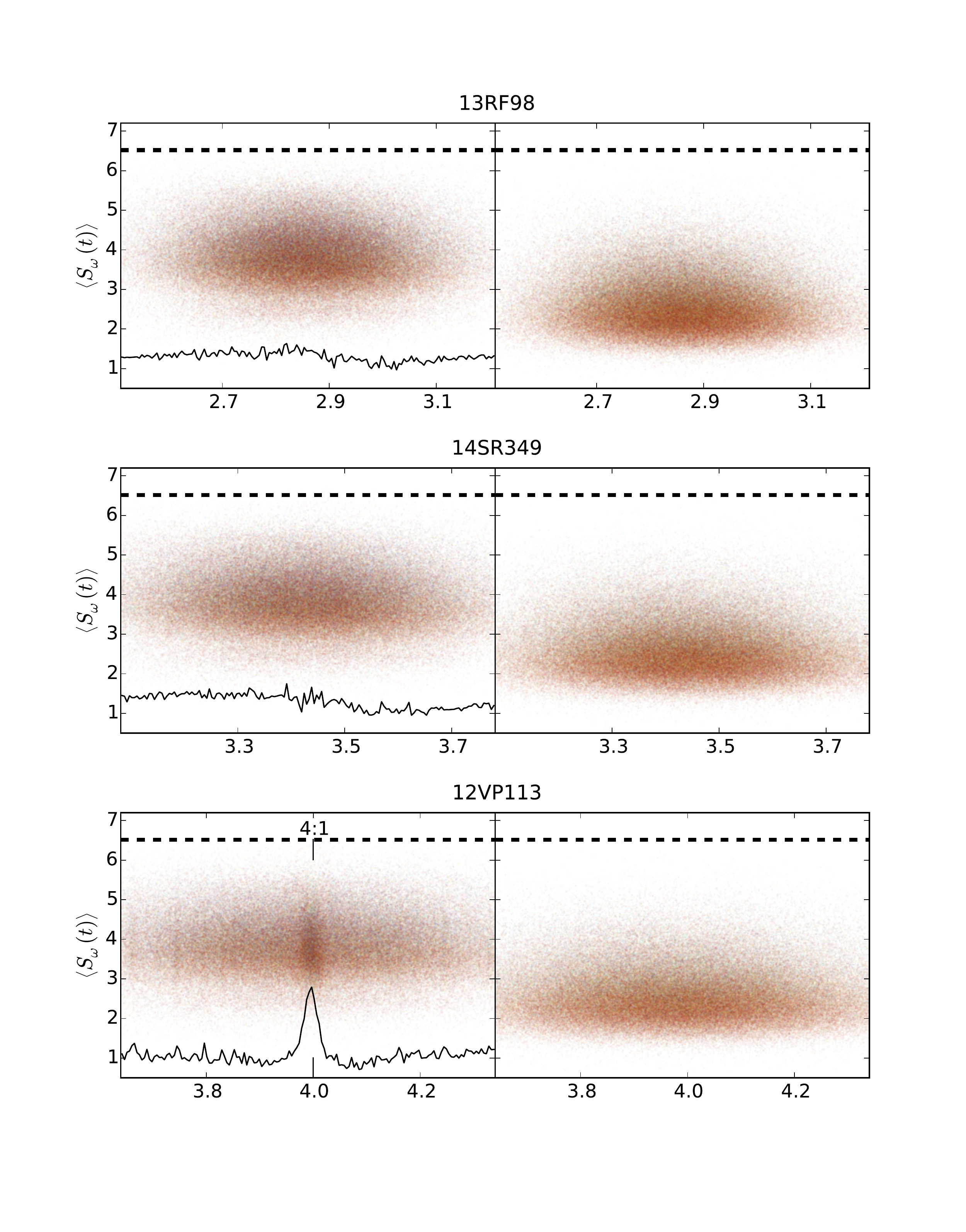}
\caption{The remaining set of seven KBOs in analogous plots to Figure~\ref{period_ratio1}.}
\label{period_ratio2}
\end{centering}
\end{figure}

If mean-motion resonances are responsible for the observed apsidal clustering of the KBOs, then the configurations that exhibit better maintenance of clustering than others should preferentially place the bodies in period ratio commensurabilities. We investigated this hypothesis in the following manner. For each of the $\sim$ 250,000 integrations that was performed, we extracted the single multiplexed sample combination of seven KBOs with the largest value of $\left<{S_{\omega}(t)}\right>$. We also extracted one random combination of KBOs from the multiplexed samples. In the left panels of Figures ~\ref{period_ratio1} and ~\ref{period_ratio2}, we plot the clustering merit $\left<{S_{\omega}(t)}\right>$ versus the period ratio $P_{P9}/P_{KBO}$ at $t=0$ for the best-performing multiplexed samples. The right panels show the analogous plots for the random multiplexed samples.

The diagrams confirm that mean-motion resonances are in fact an influential mechanism in the observed KBO clustering. The vertical overdensities in the left panels appear at distinct period ratios, signaling that the configurations that perform the best are preferentially near 
these commensurabilties. We have labeled the commensurabilities that manifest most strongly with their integer ratios. To further guide the eye, differenced histograms of the period ratios are plotted as black lines in the left panels. The differenced histograms are made by generating histograms of the period ratios of the best-performing samples (those in the left panels) with $\left<{S_{\omega}(t)}\right> > 4$ and subtracting from them Gaussian fits to the period ratio histograms of the random samples (right panels). The peaks in the resulting differenced histograms indicate the locations of the overdensities.

2012 VP\textsubscript{113} exhibits the strongest feature at 4:1, with significant features also in 2010 GB\textsubscript{174} at 9:4, 7:3, and 5:2, and in Sedna at 3:2. Interestingly, Sedna displays several other overdense striations at nearby, higher order resonances.

Although these period ratio diagrams establish the importance of proximity to orbital period commensurabilities, we have not yet demonstrated true resonant libration. If the KBOs are in resonance with Planet Nine, their motions will display librations of a critical resonant angle of the form
\begin{equation}
\phi_r = (p+q)\lambda_{P9} - p\lambda_{KBO} - (q-r)\varpi_{P9} -r\varpi_{KBO}
\end{equation}
where $\lambda$ is the mean longitude and $\varpi$ is the longitude of perihelion. $p$ and $q$ are integers that are fixed for a given $(p+q):p$ resonance, and $r$ is an integer ranging from 0 to $q$. We monitored the resonant angles of the  KBOs that appear most strongly in Figures ~\ref{period_ratio1} and ~\ref{period_ratio2}: Sedna in 3:2, 2012 VP\textsubscript{113} in 4:1, and 2010 GB\textsubscript{174} in  9:4, 7:3, and 5:2. We also monitored 2004 VN\textsubscript{112} in 3:1, 2014 SR\textsubscript{349} in 7:2, 2007 TG\textsubscript{422} in 8:5, 2001 FP\textsubscript{185} in 5:1, and 2000 CR\textsubscript{105} in 5:1. (The last two are members of the $200$ AU $< a_{KBO} < 250$ AU set of KBOs that we integrated but did not include in the $\left<{S_{\omega}(t)}\right>$ clustering measurements.) Crucially, within KBO semi-major axis uncertainties, all of these resonances can coexist with $a_{P9} \sim 654$ AU.

In order to investigate both temporary and long-lasting resonant libration, we divided the billion-year integrations into 10 sections of 100 Myr and checked for libration in each slice of time. Remarkably, all KBOs we monitored were capable of experiencing resonant libration given relatively similar Planet Nine parameters. While the existence of resonant libration should not be surprising, we note that this is the first direct numerical confirmation that these observed high-eccentricity KBOs are capable of librating stably in a mean-motion resonance with a distant perturber. 

\begin{figure}[!h]
\begin{centering}
\epsscale{1.3}
\plotone{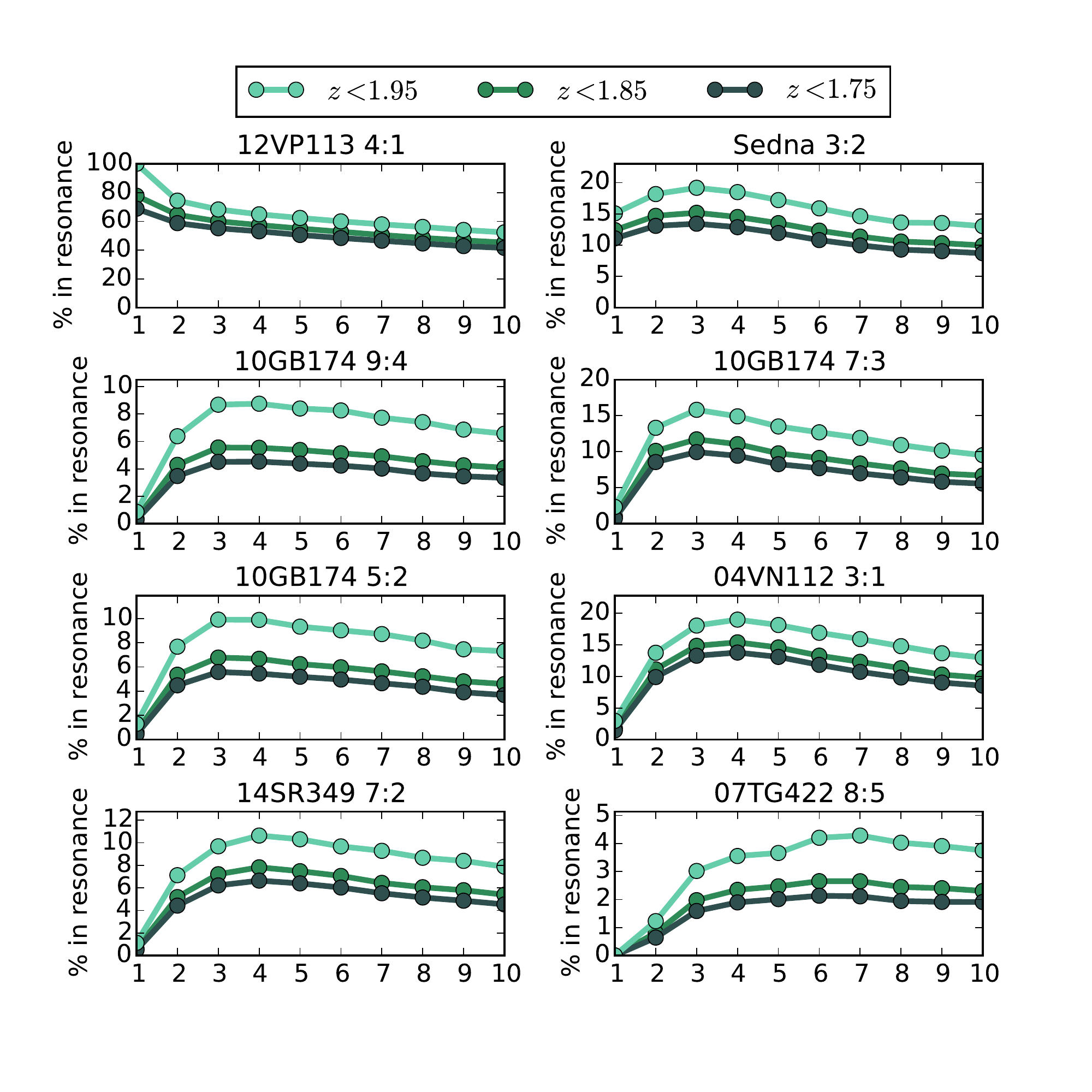}
\caption{The percentage of samples for which at least one clone of a given KBO is librating within a 100 Myr time section. The percentage is taken with respect to the sample space of configurations that produced libration with $z = \min\left\{\Delta\cos\phi_r, \Delta\sin\phi_r\right\} < 1.95$ in 2012 VP\textsubscript{113} in the first 100 Myr. (Here, $\Delta\cos\phi_r = (\cos\phi_r)_{\mathrm{max}} - (\cos\phi_r)_{\mathrm{min}}$ and likewise for $\Delta\sin\phi_r$.) The x-axis is an index with 1 representing libration in 0-100 Myr, 2 libration in 100-200 Myr, and so on. We show libration bounded by three different values of $z$, which correspond to different resonant libration amplitudes.}
\label{percentage_in_resonance}
\end{centering}
\end{figure}

 Figure~\ref{percentage_in_resonance} depicts the coexistence of resonant libration. We first restricted the sample space to the subset of Planet Nine configurations for which at least one of 2012 VP\textsubscript{113}'s critical resonant angles, $\phi_r$, was librating with $z < 1.95$ in the first 100 Myr. Here, $z \equiv \min\left\{\Delta\cos\phi_r, \Delta\sin\phi_r\right\}$, where $\Delta\cos\phi_r \equiv (\cos\phi_r)_{\mathrm{max}} - (\cos\phi_r)_{\mathrm{min}}$ and likewise for $\Delta\sin\phi_r$. For each KBO and within each 100 Myr section of time, we then calculated the percentage of samples for which at least one clone was librating during that 100 Myr slice of time. The results are shown in Figure~\ref{percentage_in_resonance}. The scales of the percentages for each KBO are relatively unimportant, as they depend on the prior distributions of $a_{P9}$ and $a_{KBO}$, but the shapes of the curves are more informative. The curves for Sedna demonstrate that it is capable of libration from the beginning of the integration, starting at 0 - 100 Myr, in a configuration in which 2012 VP\textsubscript{113} is also librating from $t=0$. The sharp increases in the percentages for most of the other resonances indicate that these bodies are difficult to maintain in resonance starting from $t=0$, but they can be captured into libration as time advances.

\begin{figure}[!tbp]
\epsscale{1.2}
\plotone{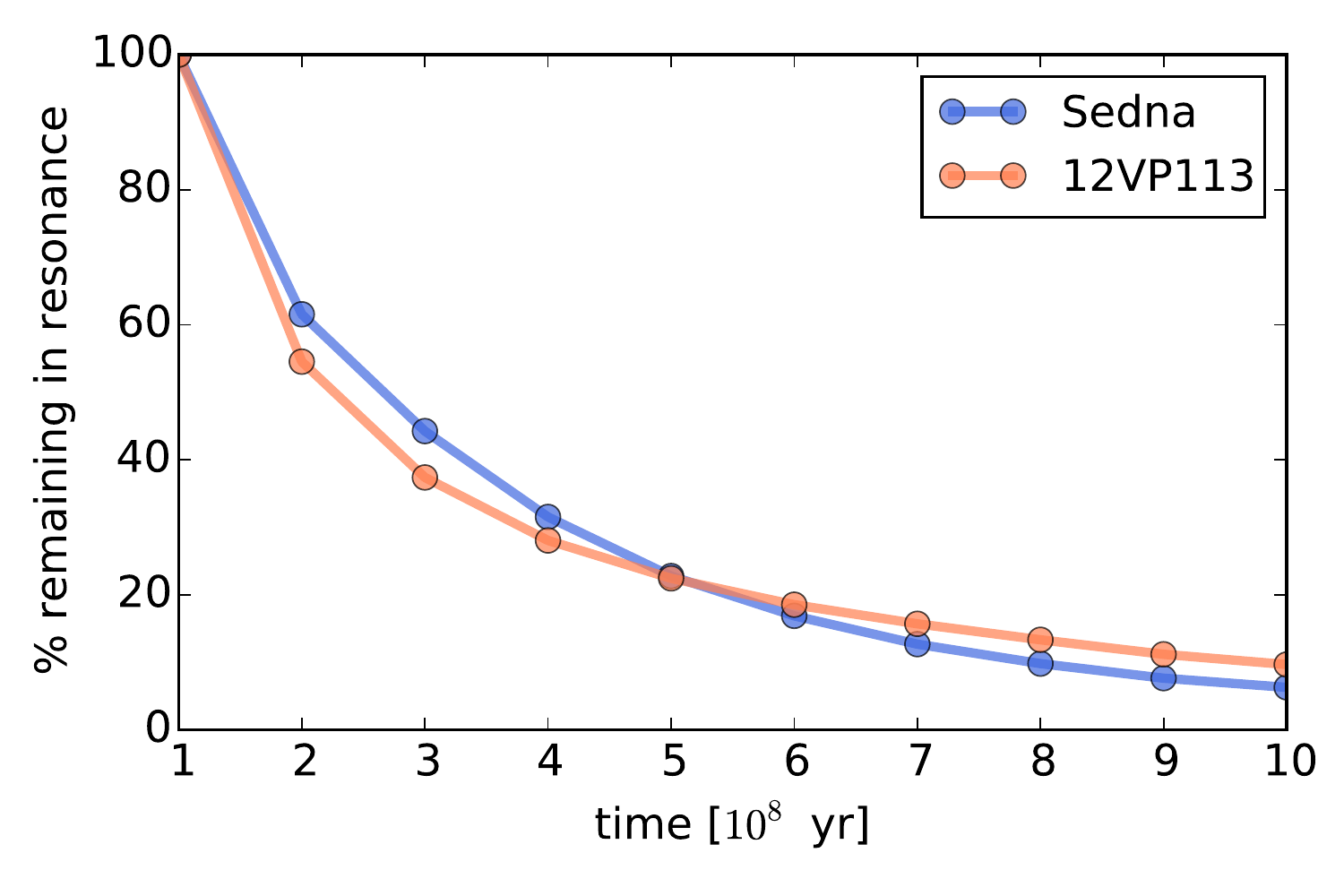}
\caption{Resonance decay for 2012 VP\textsubscript{113} in 4:1 and Sedna in 3:2. The y-axis gives the percentage of samples with at least one clone remaining in resonance starting from $t=0$ and lasting for different lengths of time. Specifically, the first point on the curve represents sustained libration for 0-100 Myr, the second for 0-200 Myr, etc. The libration amplitude cutoff is $z = \min\left\{\Delta\cos\phi_r, \Delta\sin\phi_r\right\} < 1.95$. A significant number of Planet Nine configurations keep the bodies in resonance for the full billion years. }
\label{resonance_decay}
\end{figure}

\begin{figure}[!tbp]
\epsscale{1.1}
\plotone{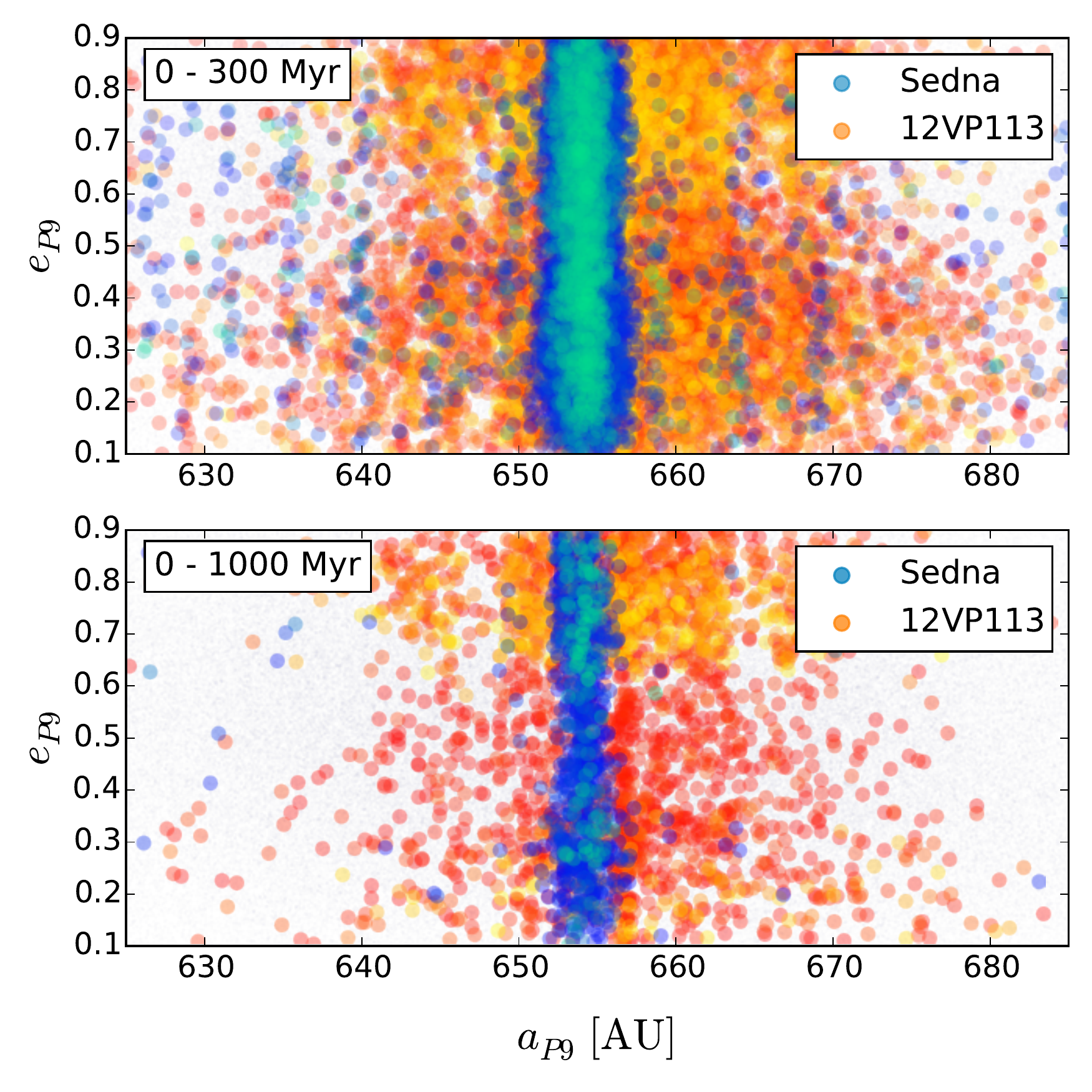}
\caption{$a_{P9}/e_{P9}$ diagrams illustrating shrinkage in the libration zones of Sedna and 2013 VP\textsubscript{113}. The spread in $a_{P9}$ reflects both the intrinsic libration amplitude and the KBO semi-major axis uncertainties. A body is considered to be in resonance if at least one of its critical resonant angles is librating, where the libration amplitude cutoff is $z = \min\left\{\Delta\cos\phi_r, \Delta\sin\phi_r\right\} < 1.95$. If more than one angle is librating, $z$ corresponds to the angle with the smallest libration amplitude. The blue/green colors indicate libration for Sedna, with greener colors indicating smaller libration amplitudes. The more orange colors for 2012 VP\textsubscript{113} are smaller amplitude.}
\label{a_vs_e_Sedna_VP}
\end{figure}

We also examined long-lived libration, that is, resonance starting from $t=0$ and lasting for an arbitrary length of time, rather than libration within a 100 Myr section of time. Figure~\ref{resonance_decay} shows the decay in the maintenance of resonant libration of Sedna and 2012 VP\textsubscript{113} as a function of time. The percentage of samples with the bodies remaining in resonance decays predictably as time advances, though there is a substantial fraction of the initial population that maintains libration for the full billion-year integration. The libration zones of Sedna and 2013 VP\textsubscript{113} are illustrated on an $a_{P9}/e_{P9}$ diagram in Figure~\ref{a_vs_e_Sedna_VP}. Here, the spread in $a_{P9}$ for the resonant configurations reflects \textit{both} the intrinsic libration width and the uncertainties propagated through the KBO semi-major axes. The libration zones shrink in time as there is a decay in the number of bodies in resonance from $t=0$. However, the overlap in the libration zones remains at values of $a_{P9}/e_{P9}$ that are consistent between the two resonances.

\section{Discussion}

\subsection{The resonant perturbation mechanism and sky location constraints}

\cite{2016AJ....151...22B} were the first to propose that mean-motion resonant perturbation may be responsible for the perihelion clustering of the distant KBOs. Subsequently, \cite{2016ApJ...824L..22M}
observed that many KBOs would exhibit small integer commensurabilities if the semi-major axis of Planet Nine was at $\sim 654$ AU. Our work has provided evidence that supports both of these claims.

The $a \sim 654$ AU resonant hypothesis will be rapidly strengthened or undermined by future KBO observations. Participation in resonant libration generates predicted bounds on the KBO semi-major axes. These bounds are substantially smaller than current measurement uncertainties such as those for 2012 VP\textsubscript{113}, which we have shown can exhibit long-lasting libration in a 4:1 resonance. Repeated observations of these KBOs to refine their period ratios can therefore permit rapid testing of the predictions, to either refute or substantially strengthen the validity of our assumed scenario. Moreover, the semi-major axes of as-yet-undiscovered, distant KBOs and their proximity to commensurabilities with $a \sim 654$ AU will continue to test the claim.

The mean-motion resonances impose restrictions not only on the orbit of the perturbing planet, but also on its position within the orbit. We can therefore evaluate the constraints our results yield on the sky position of Planet Nine, and compare the results with those of other studies. 

\begin{figure}
\epsscale{1.25}
\plotone{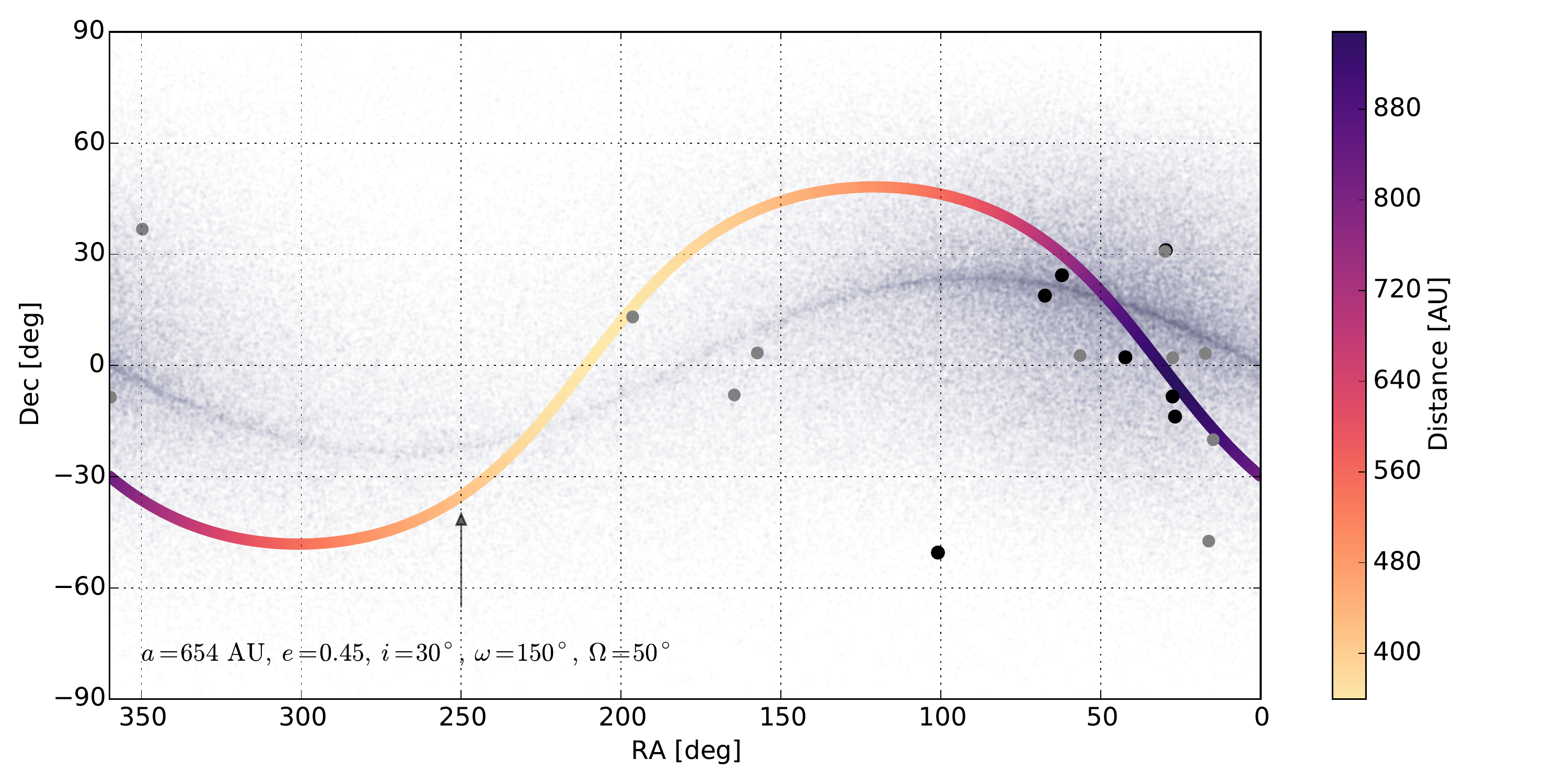}
\caption{Potential sky locations for Planet Nine. Displayed with large black points, we show the sky positions of samples with strong anti-alignment, $\left<{A_{\omega}(t)}\right> < -0.97$, and strong clustering, $\left<{S_{\omega}(t)}\right> > 5.2$. The large gray points are the sky positions of resonant configurations subject to the following constraints: Sedna and 2012 VP\textsubscript{113} librating from $t=0$ for at least 100 Myr, 2010 GB\textsubscript{174} librating in one of its three dominant resonances, and $0.4 < e_{P9} < 0.5$. The gray faint background displays all samples. The line corresponds to our fiducial orbit with orbital elements listed in the lower left, and the coloration is according to distance. (Note that the coloration is not a one-to-one correspondence with Figure~\ref{3D_orbits}.) }
\label{RA_Dec}
\end{figure}

In Figure~\ref{RA_Dec}, we display with large black points the sky positions of configurations where the KBOs maintained strong anti-alignment with Planet Nine, $\left<{A_{\omega}(t)}\right> < -0.97$, and strong clustering in the arguments of perihelion, $\left<{S_{\omega}(t)}\right> > 5.2$. The large gray points correspond to resonant configurations with the following constraints: Sedna and 2012 VP\textsubscript{113} librating from $t=0$ for at least 100 Myr, 2010 GB\textsubscript{174} librating in one of its three dominant resonances, and $0.4 < e_{P9} < 0.5$. The faint gray points are the sky locations of all samples, which is not a uniform distribution in RA due to the orientation of the orbit in space from the restrictions on $\varpi$. Finally, the orbit of our fiducial model is shown with coloration according to distance.\footnote{A variety of Planet Nine orbits relative to the fiducial model are possible, and no orbit with representative round-numbered orbital elements is expected to be \textit{the} perfect solution. When the fiducial model is integrated for 4 billion years with cloned copies of the KBOs that are consistent with observed uncertainties, the majority of the bodies maintain orbital stability and low inclinations relative to the ecliptic. Many of the bodies also maintain apsidal clustering, with the best multiplexed sample's 1-Gyr averaged values of $\left<{S_{\omega}(t)}\right>$ and $\left<{S_{\varpi}(t)}\right>$ being 5.56 and 5.24, respectively. In general, the results are better than the outcomes of similar Planet Nine orbits that we tested for 4 Gyr.} Though the plot is sparse, it is clear that the region with the highest density of favored points occurs at RA $\sim 30^{\circ}$ and Dec $\sim -10^{\circ}$, or near aphelion of our fiducial orbit.

Furthermore, analytic considerations of the KBO resonances also point towards aphelion as a likely location. In our integrations, we found that resonant configurations were the most stable and long-lasting when the critical resonant angle $\phi_q = (p+q)\lambda_{P9} - p\lambda_{KBO} - q\varpi_{KBO}$ exhibited libration about centers at $\phi_q=0^{\circ}$ or $\phi_q = 180^{\circ}$. We calculated histograms of $\phi_q$ libration amplitudes exhibited by the KBOs when they were in resonance. Then, under the assumption of libration about $\phi_q = 0^{\circ}$ or $\phi_q = 180^{\circ}$, we transformed these observed distributions of KBO libration amplitudes into a joint probability distribution of the present-day mean anomaly for Planet Nine. The resulting most likely locations are perihelion or aphelion, which may be understood as a symmetry resulting from the fact that the KBOs are all currently near their perihelion passages. Moreover, observational constraints significantly disfavor perihelion for Planet Nine, making aphelion the most probable location. 
 
\begin{figure*}[!tbp]
\begin{centering}
\epsscale{0.9}
\plotone{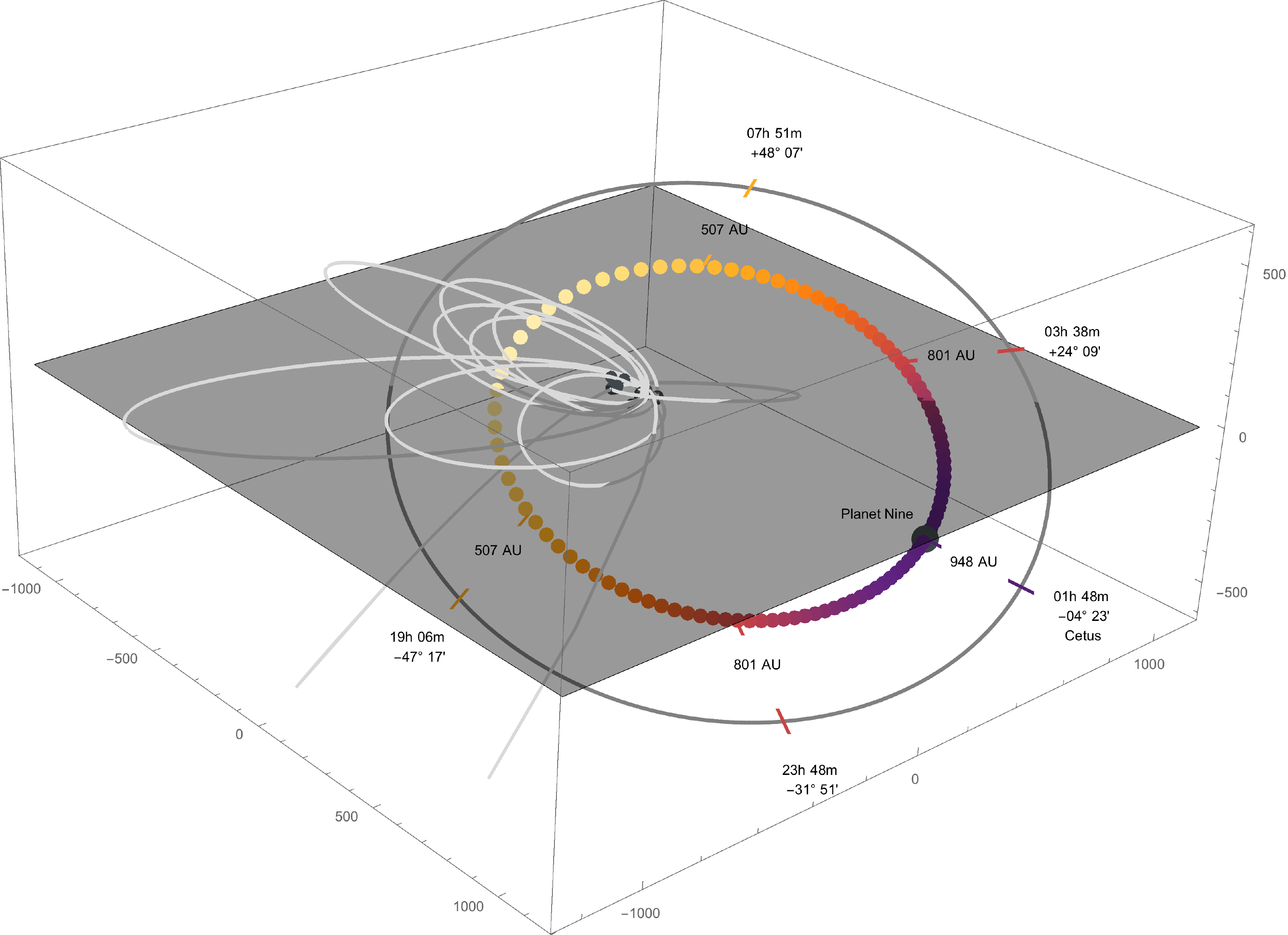}
\caption{A 3D plot of the orbits of the distant KBOs and our fiducial Planet Nine model. The transparent gray plane is the ecliptic. The orbits in white correspond to all KBOs with $a > 250$ AU, including the oppositely-aligned 2013 FT\textsubscript{28} and the long-period 2004 FE\textsubscript{72} and uo3L91, which were not included in the simulations. The gray dots on the KBO orbits are the objects' current positions. The colored orbit is the fiducial model with $a = 654$ AU, $e = 0.45$, $i = 30^{\circ}$, $\omega = 150^{\circ}$, and $\Omega = 50^{\circ}$. Planet Nine has been placed at aphelion, its favored location. The points along the orbit are colored according to distance and are evenly spaced in time. (Note that the coloration is not a one-to-one correspondence with Figure~\ref{RA_Dec}.) We have labeled the distances and RA/declination coordinates of five selected locations along the orbit. A manipulable version of the 3D plot is available at the following URL: \url{https://smillholland.github.io/P9_Orbit/}. }
\label{3D_orbits}
\end{centering}
\end{figure*}

 If the planet has the size and the reflectivity of Neptune, our fiducial model suggests that it will have V$\sim23$ at aphelion, with a parallactic motion of 0.5$^{\prime\prime}~{\rm hr}^{-1}$. To visualize the favored location more concretely, a 3D plot of the KBO orbits and our fiducial Planet Nine model is shown in Figure~\ref{3D_orbits}. A manipulable version of the 3D figure is available at \url{https://smillholland.github.io/P9_Orbit/}.
 
\begin{comment} 
\begin{figure}
\epsscale{1.1}
\plotone{3D_orbits.pdf}
\caption{A 3D plot of the orbits of the distant KBOs and our fiducial Planet Nine model. The transparent gray plane is the ecliptic. The orbits in white correspond to all KBOs with $a > 250$ AU, including the oppositely-aligned 2013 FT\textsubscript{28} and the long-period 2004 FE\textsubscript{72} and uo3L91, which were not included in the simulations. The gray dots on the KBO orbits are the objects' current positions. The colored orbit is the fiducial model with $a = 654$ AU, $e = 0.45$, $i = 30^{\circ}$, $\omega = 150^{\circ}$, and $\Omega = 50^{\circ}$. Planet Nine has been placed at aphelion, its favored location. The points along the orbit are colored according to distance and are evenly spaced in time. (Note that the coloration is not a one-to-one correspondence with Figure~\ref{RA_Dec}.) We have labeled the distances and RA/declination coordinates of five selected locations along the orbit. A manipulable version of the 3D plot is available at the following URL: \url{https://smillholland.github.io/P9_Orbit/}. }
\label{3D_orbits}
\end{figure}
\end{comment}

Our current estimate of Planet Nine's sky location is consistent with previously derived constraints. \cite{2016ApJ...824L..23B} used the observational limits of previous and present sky surveys to delineate regions of phase-space where Planet Nine should have been or will be detected. Figure 10 of their paper shows that the zone of RA $\sim 30^{\circ}-90^{\circ}$ corresponds to a largely unsurveyed region on the sky. \cite{2016A_A...587L...8F} considered the nominal \cite{2016AJ....151...22B} Planet Nine with a range of orbital positions. They performed fits of the decade-long \textit{Cassini} measurements of the Earth-Saturn distance to constrain the current location of Planet Nine. They rule out Planet Nine positions near perihelion, given by RA $< 2^{\circ}$ and RA $> 255^{\circ}$, that worsen the fit to the \textit{Cassini} residuals. The orbital position that leads to a maximum improvement of the residuals is located at RA $\sim 30^{\circ}$ and Dec $\sim -20^{\circ}$, which is near our favored region. \cite{2016AJ....152...94H} used a dynamical model of the tidal influence of Planet Nine to fit the \textit{Cassini} ranging data. They significantly rule out large areas of the sky and find a most favored location of RA $\sim 40^{\circ}$ and Dec $\sim -15^{\circ}$, extending $\sim 20^{\circ}$ in all directions. Although these results depend sensitively on the assumed solar system model \citep{2016DPS....4812007F}, it is perhaps worth noting that the region correlates well with the highest density of favored points in Figure~\ref{RA_Dec}.

The sparseness of the resonant configurations in Figure~\ref{RA_Dec} is a reflection of the computational infeasibility of using $N$-body integrations alone to pinpoint the perturber parameters that allow multiple KBOs to coexist in resonance. If the resonant hypothesis is correct, an optimization scheme that more heavily exploits the MMRs will effectively constrain Planet Nine's orbital elements and sky position. Specifically, we are interested in converging upon the  parameters that allow all KBOs to exhibit long-lasting resonant libration simultaneously. This will require not just orbital integrations but also a better dynamical understanding of how these high eccentricity resonances are arranged in phase-space. 

\subsection{An alternate interpretation of the results}

The orbital element constraints in section ~\ref{orb_elmt_constraints} were presented as constraints on Planet Nine conditional upon its existence. We now release this assumption and comment on the implications of the results for the evidence of the planet's existence. In this paradigm, there is opportunity for a much less optimistic interpretation.

As shown in Figure~\ref{ecc_anti_align_and_long}, for example, the time-averaged measure of clustering in the KBO arguments of perihelion, $\left<{S_{\omega}(t)}\right>$, is almost always smaller than the present day value, ${S_{\omega}(0)} = 6.52$. In other words, sustained clustering that matches the present-day value is exceptionally difficult to obtain. The fact that it is so challenging could be viewed as evidence against the planet's existence.

One possible explanation of the discrepancy is that $\left<{S_{\omega}(t)}\right>$ might not be a suitable comparison measurement to ${S_{\omega}(0)}$. This time-averaged quantity favors clustering throughout the entire integration, even though the bodies experience perturbations in their perihelion distances that would render them unobservable to present-day surveys. Moreover, it might be the case that these high-$a$, high-$e$ KBOs are more clustered than as-yet unobserved counterparts with similarly large $a$ but smaller $e$. If that is the case, a variant of $\left<{S_{\omega}(t)}\right>$ that is restricted to bodies with perihelion distances below some limit of observability would be a more suitable comparison metric to ${S_{\omega}(0)}$. However, we checked the integrations for a connection between observability (i.e. small perihelion distances) and clustering in $\omega$ and $\varpi$ and found little evidence for such a connection. There is thus no reason to expect that $\left<{S_{\omega}(t)}\right>$ is a biased merit function, so the tension between $\left<{S_{\omega}(t)}\right>$ and ${S_{\omega}(0)}$ remains.  

One way to reconcile the difficulty the KBOs have in maintaining clustering is to take it as evidence against the existence of Planet Nine. An alternate, more optimistic interpretation involves two ideas: 1) we might be observing the KBOs at a time period in which they are more clustered than average, such that $\left<{S_{\omega}(t)}\right>$ need not match ${S_{\omega}(0)}$; and 2) it is possible that only a small but nonzero domain of Planet Nine parameters are capable of producing long-lasting clustering in the observed KBOs. \\

\subsection{In conclusion}

The prospect that our solar system harbors an as-yet undetected super-Earth mass planet has generated an extraordinary wave of interest and attention. Two facets of the hypothesis seem particularly compelling. If the planet does exist, and especially if its mass is less than $10 M_{\oplus}$, it can provide a detailed ground-truth verification that theoretical models of the atmospheres and structures of planets in unexplored mass and temperature regimes are on the right track. And there is a satisfyingly dramatic prospect of resolution. If a trans-Neptunian planet with the proposed properties is \textit{actually} out there, it will detected very soon.

\section{Acknowledgements} 

We are thankful to the referee Alessandro Morbidelli for providing a thorough review that greatly improved the manuscript. We are also grateful to Konstantin Batygin for discussions and extensive comments on the paper. We are thankful to Hanno Rein for an inspiring conversation. Finally, we also acknowledge the Yale Center for Research Computing for use of its high performance computing clusters, and we especially thank Daisuke Nagai, Steve Weston, Kaylea Nelson, and Andrew Sherman for their assistance. This material is based upon work supported by the National Aeronautics and Space Administration through the NASA Astrobiology Institute under Cooperative Agreement Notice NNH13ZDA017C issued through the Science Mission Directorate. We acknowledge support from the NASA Astrobiology Institute through a cooperative agreement between NASA Ames Research Center and Yale University.

\bibliographystyle{apj}
\bibliography{Planet9_submitted_v1_arXiv}

\begin{thebibliography}{}
%\expandafter\ifx\csname natexlab\endcsname\relax\def\natexlab#1{#1}\fi

\bibitem[{{Bailey} {et~al.}(2016){Bailey}, {Batygin}, \&
  {Brown}}]{2016AJ....152..126B}
{Bailey}, E., {Batygin}, K., \& {Brown}, M.~E. 2016, \aj, 152, 126

\bibitem[{{Batygin} \& {Brown}(2016{\natexlab{a}})}]{2016AJ....151...22B}
{Batygin}, K., \& {Brown}, M.~E. 2016{\natexlab{a}}, \aj, 151, 22

\bibitem[{{Batygin} \& {Brown}(2016{\natexlab{b}})}]{2016ApJ...833L...3B}
---. 2016{\natexlab{b}}, \apjl, 833, L3

\bibitem[{{Beust}(2016)}]{2016A_A...590L...2B}
{Beust}, H. 2016, \aap, 590, L2

\bibitem[{{Bromley} \& {Kenyon}(2016)}]{2016ApJ...826...64B}
{Bromley}, B.~C., \& {Kenyon}, S.~J. 2016, \apj, 826, 64

\bibitem[{{Brown} \& {Batygin}(2016)}]{2016ApJ...824L..23B}
{Brown}, M.~E., \& {Batygin}, K. 2016, \apjl, 824, L23

\bibitem[{{Burns}(1976)}]{Burns1976}
{Burns}, J.~A. 1976, AmJPh, 44, 944

\bibitem[{{Chambers}(1999)}]{1999MNRAS.304..793C}
{Chambers}, J.~E. 1999, \mnras, 304, 793

\bibitem[{{Chen} {et~al.}(2016){Chen}, {Lin}, {Holman}, {Payne}, {Fraser},
  {Lacerda}, {Ip}, {Chen}, {Kudritzki}, {Jedicke}, {Wainscoat}, {Tonry},
  {Magnier}, {Waters}, {Kaiser}, {Wang}, \& {Lehner}}]{2016ApJ...827L..24C}
{Chen}, Y.-T., {Lin}, H.~W., {Holman}, M.~J., {et~al.} 2016, \apjl, 827, L24

\bibitem[{{de la Fuente Marcos} \& {de la Fuente
  Marcos}(2014)}]{2014MNRAS.443L..59D}
{de la Fuente Marcos}, C., \& {de la Fuente Marcos}, R. 2014, \mnras, 443, L59

\bibitem[{{de la Fuente Marcos} {et~al.}(2016){de la Fuente Marcos}, {de la
  Fuente Marcos}, \& {Aarseth}}]{2016MNRAS.460L.123D}
{de la Fuente Marcos}, C., {de la Fuente Marcos}, R., \& {Aarseth}, S.~J. 2016,
  \mnras, 460, L123

\bibitem[{{Fienga} {et~al.}(2016){Fienga}, {Laskar}, {Manche}, \&
  {Gastineau}}]{2016A_A...587L...8F}
{Fienga}, A., {Laskar}, J., {Manche}, H., \& {Gastineau}, M. 2016, \aap, 587,
  L8

\bibitem[{{Folkner} {et~al.}(2016){Folkner}, {Jacobson}, {Park}, \&
  {Williams}}]{2016DPS....4812007F}
{Folkner}, W., {Jacobson}, R.~A., {Park}, R., \& {Williams}, J.~G. 2016, in
  AAS/Division for Planetary Sciences Meeting Abstracts, Vol.~48, AAS/Division
  for Planetary Sciences Meeting Abstracts

\bibitem[{{Fortney} {et~al.}(2016){Fortney}, {Marley}, {Laughlin},
  {Nettelmann}, {Morley}, {Lupu}, {Visscher}, {Jeremic}, {Khadder}, \&
  {Hargrave}}]{2016ApJ...824L..25F}
{Fortney}, J.~J., {Marley}, M.~S., {Laughlin}, G., {et~al.} 2016, \apjl, 824,
  L25

\bibitem[{{Gomes} {et~al.}(2016){Gomes}, {Deienno}, \&
  {Morbidelli}}]{2016arXiv160705111G}
{Gomes}, R., {Deienno}, R., \& {Morbidelli}, A. 2016, ArXiv e-prints,
  arXiv:1607.05111

\bibitem[{{Gomes} {et~al.}(2015){Gomes}, {Soares}, \&
  {Brasser}}]{2015Icar..258...37G}
{Gomes}, R.~S., {Soares}, J.~S., \& {Brasser}, R. 2015, \icarus, 258, 37

\bibitem[{{Holman} \& {Payne}(2016{\natexlab{a}})}]{2016AJ....152...80H}
{Holman}, M.~J., \& {Payne}, M.~J. 2016{\natexlab{a}}, \aj, 152, 80

\bibitem[{{Holman} \& {Payne}(2016{\natexlab{b}})}]{2016AJ....152...94H}
---. 2016{\natexlab{b}}, \aj, 152, 94

\bibitem[{{Hoyt}(1980)}]{1980pxp..book.....H}
{Hoyt}, W.~G. 1980, {Planets X and Pluto}

\bibitem[{{Kenyon} \& {Bromley}(2016)}]{2016ApJ...825...33K}
{Kenyon}, S.~J., \& {Bromley}, B.~C. 2016, \apj, 825, 33

\bibitem[{{Lai}(2016)}]{2016arXiv160801421L}
{Lai}, D. 2016, ArXiv e-prints, arXiv:1608.01421

\bibitem[{{Li} \& {Adams}(2016)}]{2016ApJ...823L...3L}
{Li}, G., \& {Adams}, F.~C. 2016, \apjl, 823, L3

\bibitem[{{Linder} \& {Mordasini}(2016)}]{2016A&A...589A.134L}
{Linder}, E.~F., \& {Mordasini}, C. 2016, \aap, 589, A134

\bibitem[{{Malhotra} {et~al.}(2016){Malhotra}, {Volk}, \&
  {Wang}}]{2016ApJ...824L..22M}
{Malhotra}, R., {Volk}, K., \& {Wang}, X. 2016, \apjl, 824, L22

\bibitem[{{Mustill} {et~al.}(2016){Mustill}, {Raymond}, \&
  {Davies}}]{2016MNRAS.460L.109M}
{Mustill}, A.~J., {Raymond}, S.~N., \& {Davies}, M.~B. 2016, \mnras, 460, L109

\bibitem[{{Shankman} {et~al.}(2016){Shankman}, {Kavelaars}, {Lawler},
  {Gladman}, \& {Bannister}}]{2016arXiv161004251S}
{Shankman}, C., {Kavelaars}, J., {Lawler}, S.~M., {Gladman}, B.~J., \&
  {Bannister}, M.~T. 2016, ArXiv e-prints, arXiv:1610.04251

\bibitem[{{Sheppard} \& {Trujillo}(2016)}]{2016AJ....152..221S}
{Sheppard}, S.~S., \& {Trujillo}, C. 2016, \aj, 152, 221

\bibitem[{{Ter Braak}(2006)}]{TerBraak2006}
{Ter Braak}, C. J.~F. 2006, Stat. Comput., 16, 239

\bibitem[{{Toth}(2016)}]{2016A&A...592A..86T}
{Toth}, I. 2016, \aap, 592, A86

\bibitem[{{Trujillo} \& {Sheppard}(2014)}]{2014Natur.507..471T}
{Trujillo}, C.~A., \& {Sheppard}, S.~S. 2014, \nat, 507, 471


\end{thebibliography}

\begin{comment}

\end{comment}

\end{document}